\documentclass[twocolumn,aps,prl,superscriptaddress,amsmath,amssym,floatfix,nofootinbib]{revtex4-1}

\usepackage{graphicx}
\usepackage{bm}
\usepackage{color}
\usepackage[usenames,dvipsnames]{xcolor}
\usepackage{tikz}
\usetikzlibrary{%
    decorations.pathreplacing,%
    decorations.pathmorphing,%
    arrows
}
\usepackage{amssymb}
\usepackage{enumerate}
\usepackage{comment}

\def\be{\begin{equation}}
\def\ee{\end{equation}}

\newcommand{\ket}[1]{\left\vert #1 \right\rangle}
\newcommand{\bra}[1]{\left\langle #1 \right\vert}
\newcommand{\ip}[2]{\langle #1 \vert #2 \rangle}

\usepackage[colorlinks=true,linkcolor=MidnightBlue,citecolor=blue,filecolor=green,urlcolor=PineGreen]{hyperref}

\begin{document}
\title{
Diffusive entanglement generation by \\continuous homodyne monitoring of spontaneous emission
}
\author{Philippe Lewalle} 
\email{plewalle@ur.rochester.edu}
\affiliation{Department of Physics and Astronomy, University of Rochester, Rochester, NY 14627, USA}
\affiliation{Center for Coherence and Quantum Optics, University of Rochester, Rochester, NY 14627, USA}
\author{Cyril Elouard} 
\affiliation{Department of Physics and Astronomy, University of Rochester, Rochester, NY 14627, USA}
\affiliation{Center for Coherence and Quantum Optics, University of Rochester, Rochester, NY 14627, USA}
\author{Sreenath K. Manikandan} 
\affiliation{Department of Physics and Astronomy, University of Rochester, Rochester, NY 14627, USA}
\affiliation{Center for Coherence and Quantum Optics, University of Rochester, Rochester, NY 14627, USA}
\author{Xiao--Feng Qian} 
\affiliation{Department of Physics and Astronomy, University of Rochester, Rochester, NY 14627, USA}
\affiliation{Center for Coherence and Quantum Optics, University of Rochester, Rochester, NY 14627, USA}
\affiliation{Department of Physics and Center for Quantum Science and Engineering, Stevens Institute of Technology, Hoboken, NJ 07030, USA}
\author{Joseph H. Eberly} 
\affiliation{Department of Physics and Astronomy, University of Rochester, Rochester, NY 14627, USA}
\affiliation{Center for Coherence and Quantum Optics, University of Rochester, Rochester, NY 14627, USA}
\author{Andrew N. Jordan} 
\affiliation{Department of Physics and Astronomy, University of Rochester, Rochester, NY 14627, USA}
\affiliation{Center for Coherence and Quantum Optics, University of Rochester, Rochester, NY 14627, USA}
\affiliation{Institute for Quantum Studies, Chapman University, Orange, CA 92866, USA}
\date{\today}

\begin{abstract}
We consider protocols to generate quantum entanglement between two remote qubits, through joint time--continuous detection of their spontaneous emission. We demonstrate that schemes based on homodyne detection, leading to diffusive quantum trajectories, lead to identical average entanglement yield as comparable photodetection strategies; this is despite substantial differences in the two--qubit state dynamics between these schemes, which we explore in detail. The ability to use different measurements to achieve the same ends may be of practical significance; the less--well--known diffusive scheme appears far more feasible on superconducting qubit platforms in the near term.

\end{abstract}

\maketitle

\par
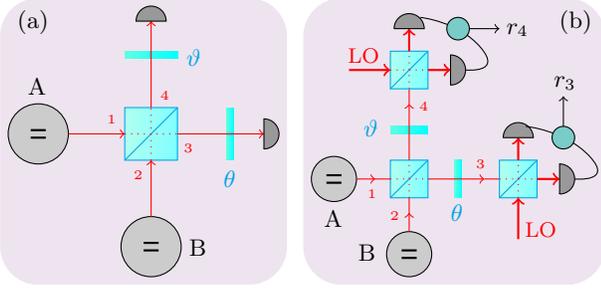
\begin{figure}
\centering \begin{tikzpicture}
    \filldraw[fill = Plum!10!white, draw = Plum!10!white, rounded corners = 0.5cm] (0,0) rectangle (3.8,3.8);
    \filldraw[fill=black!20!white, draw=black] (0.5,2) circle (0.4 cm);
    \filldraw[fill=black!20!white, draw=black] (2,0.5) circle (0.4 cm);
    \draw[thick] (0.4,2.05) -- (0.6,2.05);
    \draw[thick] (0.4,1.95) -- (0.6,1.95);
    \draw[thick] (1.9,0.55) -- (2.1,0.55);
    \draw[thick] (1.9,0.45) -- (2.1,0.45);
    \shade[left color=SkyBlue!50,right color=Cyan!50] (1.65,1.65) rectangle (2.35,2.35);
    \draw[Cerulean] (1.65,1.65) -- (1.65,2.35) -- (2.35,2.35) -- (2.35,1.65) -- cycle;
    \draw[Cerulean] (1.65,1.65) -- (2.35,2.35);
    \draw[red,->] (0.9,2) -- (1.65,2);
    \draw[red,->] (2,0.9) -- (2,1.65);
    \draw[red,dotted] (1.65,2) -- (2.35,2);
    \draw[red,dotted] (2,1.65) -- (2,2.35);
    \shade[left color=SkyBlue,right color=Cyan] (3,1.65) rectangle (3.1,2.35);
    \shade[left color=SkyBlue,right color=Cyan] (1.65,3) rectangle (2.35,3.1);
    \draw[red,->] (2,2.35) -- (2,3.5);
    \draw[red,->] (2.35,2) -- (3.5,2);
    \filldraw[fill=black!40!white] (3.5,1.8) -- (3.5,2.2) arc (90:-90:0.2cm) -- (3.5,1.8);
    \filldraw[fill=black!40!white] (1.8,3.5) -- (2.2,3.5) arc (0:180:0.2cm) -- (1.8,3.5);
\end{tikzpicture} \hspace{0pt} \begin{tikzpicture}
    \filldraw[fill = Plum!10!white, draw = Plum!10!white, rounded corners = 0.5cm] (-0.1,0) rectangle (3.9,3.8);
    \filldraw[fill=black!20!white, draw=black] (0.3,1.4) circle (0.3 cm);
    \filldraw[fill=black!20!white, draw=black] (1.3,0.4) circle (0.3 cm);
    \draw[thick,black] (0.2,1.35) -- (0.4,1.35);
    \draw[thick,black] (0.2,1.45) -- (0.4,1.45);
    \draw[thick,black] (1.2,0.35) -- (1.4,0.35);
    \draw[thick,black] (1.2,0.45) -- (1.4,0.45);
    \shade[left color=SkyBlue!50,right color=Cyan!50] (1.05,1.15) rectangle (1.55,1.65);
    \draw[Cerulean] (1.05,1.15) -- (1.05,1.65) -- (1.55,1.65) -- (1.55,1.15) -- cycle;
    \draw[Cerulean] (1.05,1.15) -- (1.55,1.65);
    \shade[left color=SkyBlue,right color=Cyan] (1.05,2.0) rectangle (1.55,2.1);
    \shade[left color=SkyBlue,right color=Cyan] (1.9,1.15) rectangle (2.0,1.65);
    \draw[red,->] (0.6,1.4) -- (0.85,1.4);
    \draw[red] (0.85,1.4) -- (1.05,1.4);
    \draw[red,->] (1.3,0.7) -- (1.3,0.95);
    \draw[red] (1.3,0.95) -- (1.3,1.15);
    \draw[red,dotted] (1.05,1.4) -- (1.55,1.4);
    \draw[red,dotted] (1.3,1.15) -- (1.3,1.65);
    \draw[red,->] (1.55,1.4) -- (2.3,1.4);
    \draw[red,->] (1.3,1.65) -- (1.3,2.4);
    \draw[red] (2.3,1.4) -- (2.5,1.4);
    \draw[red] (1.3,2.4) -- (1.3,2.6);
    \shade[left color=SkyBlue!50,right color=Cyan!50] (2.5,1.15) rectangle (3,1.65);
    \draw[Cerulean] (2.5,1.15) -- (2.5,1.65) -- (3,1.65) -- (3,1.15) -- cycle;
    \draw[Cerulean] (2.5,1.15) -- (3,1.65);
    \shade[left color=SkyBlue!50,right color=Cyan!50] (1.05,2.6) rectangle (1.55,3.1);
    \draw[Cerulean] (1.05,2.6) -- (1.05,3.1) -- (1.55,3.1) -- (1.55,2.6) -- cycle;
    \draw[Cerulean] (1.05,2.6) -- (1.55,3.1);
    \draw[red,thick,->] (0.5,2.85) -- (1.05,2.85);
    \draw[red,thick,->] (2.75,0.6) -- (2.75,1.15);
    \draw[red,dotted] (1.3,2.6) -- (1.3,3.1);
    \draw[red,dotted] (1.05,2.85) -- (1.55,2.85);
    \draw[red,dotted] (2.5,1.4) -- (3,1.4);
    \draw[red,dotted] (2.75,1.15) -- (2.75,1.65);
    \draw[red,thick,->] (3.0,1.4) -- (3.3,1.4);
    \draw[red,thick,->] (2.75,1.65) -- (2.75,1.95);
    \draw[red,thick,->] (1.3,3.1) -- (1.3,3.4);
    \draw[red,thick,->] (1.55,2.85) -- (1.85,2.85);
    \filldraw[fill=black!40!white] (1.1,3.4) -- (1.5,3.4) arc (0:180:0.2cm) -- (1.1,3.4);
    \filldraw[fill=black!40!white] (2.55,1.95) -- (2.95,1.95) arc (0:180:0.2cm) -- (2.55,1.95);
    \filldraw[fill=black!40!white] (3.3,1.2) -- (3.3,1.6) arc (90:-90:0.2cm) -- (3.3,1.2);
    \filldraw[fill=black!40!white] (1.85,2.65) -- (1.85,3.05) arc (90:-90:0.2cm) -- (1.85,2.65);
    \draw [black] plot [smooth, tension = 1.8] coordinates {(2.85,2.1) (3.7,1.7) (3.5,1.4)};
    \draw [black] plot [smooth, tension = 1.8] coordinates {(1.45,3.55) (2.25,3.15) (2.05,2.85)};
    \filldraw[draw=black,fill=Emerald!50!white] (3.35,1.95) circle (0.15 cm);
    \filldraw[draw=black,fill=Emerald!50!white] (1.95,3.4) circle (0.15 cm);
    \draw[black,->] (3.35,2.1) -- (3.35,2.5);
    \draw[black,->] (2.1,3.4) -- (2.5,3.4);
\end{tikzpicture} \begin{picture}(0,0)(144,8)
    \put(-85,105){(a)}
    \put(120,105){(b)}
    \put(-81,80){A}
    \put(-20,19){B}
    \put(-51,69){\tiny \color{red} 1}
    \put(-41,48){\tiny \color{red} 2}
    \put(-22,58){\tiny \color{red} 3}
    \put(-31,78){\tiny \color{red} 4}
    \put(-7,45){\color{Cerulean} $\theta$}
    \put(-21,91){\color{Cerulean} $\vartheta$}
    \put(31,30){A}
    \put(44,17){B}
    \put(46,64){\color{Cerulean} $\vartheta$}
    \put(79,32){\color{Cerulean} $\theta$}
    \put(47.5,41){\tiny \color{red} 1}
    \put(56,32.5){\tiny \color{red} 2}
    \put(88.5,52){\tiny \color{red} 3}
    \put(67,74){\tiny \color{red} 4}
    \put(100,103){$r_4$}
    \put(118,83){$r_3$}
    \put(107,26){\footnotesize \color{red} LO}
    \put(40,92){\footnotesize \color{red} LO}
\end{picture}
\caption{Sketch of apparatus: (a) Photodetection measures the photon number at outputs 3 and 4. 
(b) Balanced homodyne detection at both outputs (all beamspliters are 50/50) mixes the signal with a strong coherent state LO, and then monitors one quadrature, squeezing out the other. For simplicity across both measurement schemes, we introduce some phase plates at each beamsplitter output; we assume that all path lengths are equal, so that $\theta$ and $\vartheta$ completely characterize the phase relationships between signal beams and LOs. For the homodyne measurements, we could equivalently change the LO phases instead of the signal phases; the present convention is simply a matter of notational convenience.}
\label{fig-splitterphase}
\end{figure}

 Continuous monitoring of quantum systems, to generate stochastic quantum trajectories (SQTs), has become a widespread technique over the past decade 
\cite{BookCarmichael, BookPercival, BookGardiner, BookWiseman, BookBarchielli, Wiseman1996Review, Brun2001Teach, Jacobs2006, Chantasri2013, Chantasri2015, Korotkov2016, Gambetta2008, Murch2013, Weber2014}, due in large part to the development of quantum limited amplifiers \cite{Caves1982, Clerk2010Review, Bergeal2010, DevoretAmp} for circuit--QED experiments.
Continuous dispersive or parity measurements have been studied extensively for entanglement generation \cite{Ruskov2003-2, Trauzettel2006, Williams2008, Riste2013, Roch2014, martin2015remote, Motzoi2015, Silveri2016, Chantasri2016, Martin_2017}. A key ingredient in any of these entanglement schemes is that different two--qubit basis states are indistinguishable as per the relevant measurement outcomes, such that the qubits are entangled by the measurement. There has also been considerable recent work concerning the entanglement of spatially--separated atoms or spins by making joint photodetection measurements of their fluorescence \cite{Cabrillo1998, Plenio1999, BarrettKok, Lim2005, Moehring2007, Maunz2009, Hofmann2012, Santos2012, Slodika2013, Hanson2013Heralded, Pfaff2014, Delteil2016, Ohm2017, Araneda2018} (e.g.~similar to what we illustrate in Fig.~\ref{fig-splitterphase}(a)); such methods have been leveraged to realize loophole free Bell tests \cite{HansonLoopholeFree}, and Bell state measurements are a key ingredient in many proposed designs for quantum repeaters \cite{Borregarrd2019}, and studying the entanglement properties of pairs of fluorescing atoms generally has led to many fundamental insights \cite{YuEberly_2004, Xiao-Feng_2019}.

\par Our present interest lies at the intersection of these areas. While effective photon--number resolving detection does not yet, to our knowledge, exist in the microwave regime, there has been considerable recent progress in obtaining diffusive trajectories of a superconducting qubit's state by heterodyning or homodyning its spontaneous emission \cite{PCI-2013, Jordan2015flor, Campagne-Ibarcq2016, PCI-2016-2, Naghiloo2016flor, Mahdi2016, Tan2017, Ficheux2018, FlorTeach2019}. We here consider the extension of such schemes to joint measurements of a pair of qubits, i.e.~we focus on entanglement creation between two qubits, using time--continuous fluorescence measurements, according to the schemes diagrammed in Fig.~\ref{fig-splitterphase}. The most relevant pre--existing work on entanglement dynamics under such scenarios \cite{Carvalho2007, Viviescas2010} has led to a derivation of a diffusive measurement scheme based on homodyne detection (a particular unraveling of the stochastic master equation) which is optimal for entanglement preservation within a class of diffusive trajectory schemes. There is additional closely related work which has considered both photodetection and homodyne fluorescence detection \cite{Mascarenhas2011}, or other aspects of entanglement dynamics of quantum systems open to decay channels \cite{YuEberly_2004, Mintert2005, Mascarenhas2010, Carvalho2011}.
We will re--derive the optimal homodyne scheme of \cite{Viviescas2010} using a different formal and conceptual approach, and show that the average entanglement yield between this optimal diffusive scheme and photodetection are formally equivalent, even though there are considerable differences in the two--qubit state dynamics between these two schemes.
We are motivated to compare these approaches in large part by recent experimental progress in implementing diffusive trajectory experiments using fluorescence of superconducting qubits \cite{Campagne-Ibarcq2016, PCI-2016-2, Naghiloo2016flor, Mahdi2016, Tan2017, Ficheux2018}.

We use a model as in Refs.~\cite{Jordan2015flor, FlorTeach2019} to describe spontanteous emission and the change of qubit state. The qubit A and the cavity output mode it is coupled with, initially in vacuum, evolve after a time interval $dt$ from state $\ket{A_i} = \ket{0_1}\otimes (\zeta\ket{e} + \phi\ket{g})$ to
\be
\ket{A_f} = \sqrt{1-\epsilon} \zeta \ket{0e} + \phi \ket{0g} + \sqrt{\epsilon} \zeta \ket{1g}\label{psi_in},
\ee
for $\epsilon  = \gamma dt$,
and similarly for qubit B and associated cavity mode initially in state $\ket{B_i} =\ket{0_2}\otimes (\xi\ket{e} + \varphi\ket{g})$.
We suppose a qubit naturally emits into its cavity at rate $\gamma = 1/T_1$, and that the measurements are fast (i.e.~that $\epsilon \ll 1$, or $dt \ll T_1$ is the shortest timescale in our problem, and well--separated from the others in play\footnote{We also neglect the photon travel time between qubits and detectors; this is appropriate for small--scale experiments, i.e.~if $T_1 = 1\:\mu\mathrm{s}$ and $dt = T_1/100$ there is no problem neglecting the travel/delay time in an experiment on a table--top scale and treating our state update as applying in real time.}). The complex amplitudes $\zeta$ and $\phi$ specify an arbitrary pure initial state for the qubit in cavity A, and likewise for $\xi$ and $\varphi$ with respect to the qubit in cavity B. We can combine a pair of state update expressions like \eqref{psi_in} with tensor products, such that the corresponding two--qubit state update goes as 
\be \label{2Q_stateUp1}
\ket{\psi_{dt}} = \underbrace{\left( \begin{array}{cccc}
1- \epsilon & 0 & 0 & 0 \\
\sqrt{\epsilon(1-\epsilon)} \hat{a}_2^\dag & \sqrt{1-\epsilon} & 0 & 0 \\
\sqrt{\epsilon(1-\epsilon)} \hat{a}_1^\dag & 0 & \sqrt{1-\epsilon} & 0 \\
\epsilon \hat{a}_1^\dag \hat{a}_2^\dag & \sqrt{\epsilon} \hat{a}_1^\dag & \sqrt{\epsilon}\hat{a}_2^\dag & 1
\end{array} \right)}_{\mathcal{M}} \underbrace{\left( \begin{array}{c} \zeta \xi \\ \zeta \varphi \\ \phi \xi \\ \phi \varphi \end{array} \right)}_{\ket{\psi_0}} 
\ee
in the basis $(\ket{ee},\ket{eg},\ket{ge},\ket{gg})$,
where a vacuum state in both beamsplitter inputs $\ket{0_1 0_2}$ is assumed (not yet traced or projected out to leave only the qubit states), but not explicitly notated. Operators $\hat{a}^\dag$ create photons in the cavity output modes, e.g.~$\hat{a}_1^\dag \ket{0_1 0_2} = \ket{1_1 0_2}$. We have assumed that our two qubits have the same fluorescence rate $\gamma$ into their respective cavities for simplicity.

\par Kraus operators implementing two--qubit state updates for particular measurement schemes are obtained by doing suitable projections of $\mathcal{M}$ onto optical states, after the spatial modes 1 and 2 are mixed on the beamsplitter. We also draw two phase plates after each port of the beam splitter, which will play the role of turning knobs to select the quadrature of the light that will be measured. In the end, the input and output light modes are linked according to a unitary transformation $
\hat{a}_1^\dag = \tfrac{1}{\sqrt{2}}( \hat{a}_3^\dag e^{i\theta} + \hat{a}_4^\dag e^{i\vartheta} )$ and $\hat{a}_2^\dag = \tfrac{1}{\sqrt{2}}( \hat{a}_3^\dag e^{i\theta} - \hat{a}_4^\dag e^{i\vartheta})$. 
In practice, tuning the phases of the LOs may be easier than the signal, and is an entirely equivalent operation, as we only care about their relative phase.
As an example, photodetection could be modeled by considering a set of operators $\mathcal{M}_{nm} = \bra{n_3 m_4} \mathcal{M} \ket{0_3 0_4}$, where $\ket{n}$ and $\ket{m}$ are Fock states, and a complete set of photodetection outcomes (numbers of photons arriving at a detector) possible at each timestep are considered; such operators are POVM elements, i.e.~we have $\sum_{nm} \mathcal{M}_{nm}^\dag \mathcal{M}_{nm} = \mathbb{I}$.  
Homodyne measurement at both outputs involves interfering the signal beams with a strong coherent state local oscillator (LO), and effectively projecting signal onto a particular field quadrature (and squeezing out the other). 
This can be modeled by operators $\mathcal{M}_{34} = \bra{X_3 X_4} \mathcal{M} \ket{0_3 0_4}$, where the states $\ket{X}$ are eigenstates of a quadrature operator $\hat{X} = (\hat{a}+\hat{a}^\dag)/\sqrt{2}$ \cite{Wiseman1996Review, BookWiseman}. The particular values of $X_3$ and $X_4$ obtained in any given timestep are stochastic, creating the measurement records. 
Such a set of operators are also complete, in that $\int dX_3 \: dX_4 \: \mathcal{M}_{34}^\dag \mathcal{M}_{34} \propto \mathbb{I}$. 
The state can be updated as per \cite{BookNielsen}
\be \label{hom-stateup}
\rho(t+dt) = \frac{\mathcal{M}_{34} \rho(t) \mathcal{M}_{34}^\dag}{\text{tr}\left( \mathcal{M}_{34} \rho(t) \mathcal{M}_{34}^\dag \right)},
\ee
conditioned on the measurement outcomes $X_3$ and $X_4$, where $\rho$ is the two--qubit density matrix, and the outcomes are drawn from a distribution $\wp(X_3,X_4|\rho) = \text{tr}( \mathcal{M}_{34} \rho(t) \mathcal{M}_{34}^\dag)$ which is approximately Gaussian in both variables. 
It is possible to express these dynamics in terms of a Markovian stochastic master equation \cite{Jacobs2006}, which is equivalent to the formulation shown here to $O(dt)$. 
Additional details and derivations are provided in an extended version of the present work \cite{LongFlor2019}, and in \cite{FlorTeach2019}.

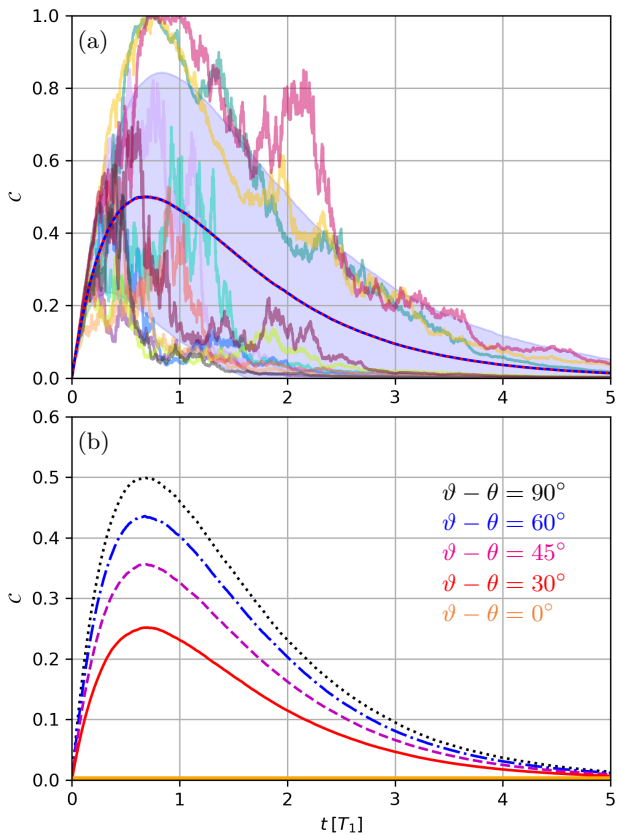
\begin{figure}
    \centering
    \includegraphics[width=\columnwidth,trim = {5pt 18pt 20pt 25pt},clip]{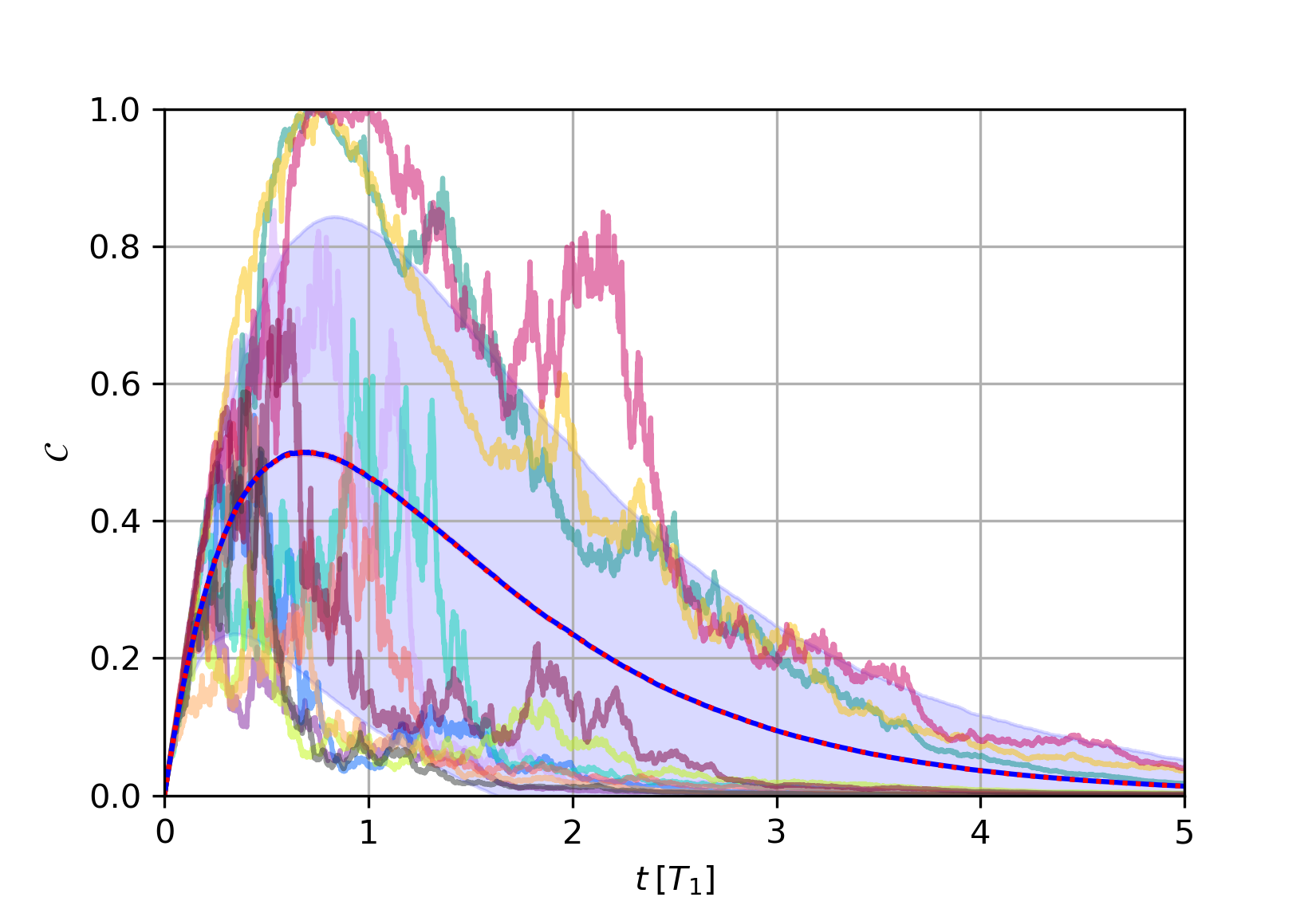} \\
    \includegraphics[width=\columnwidth,trim = {5pt 0 20pt 28pt},clip]{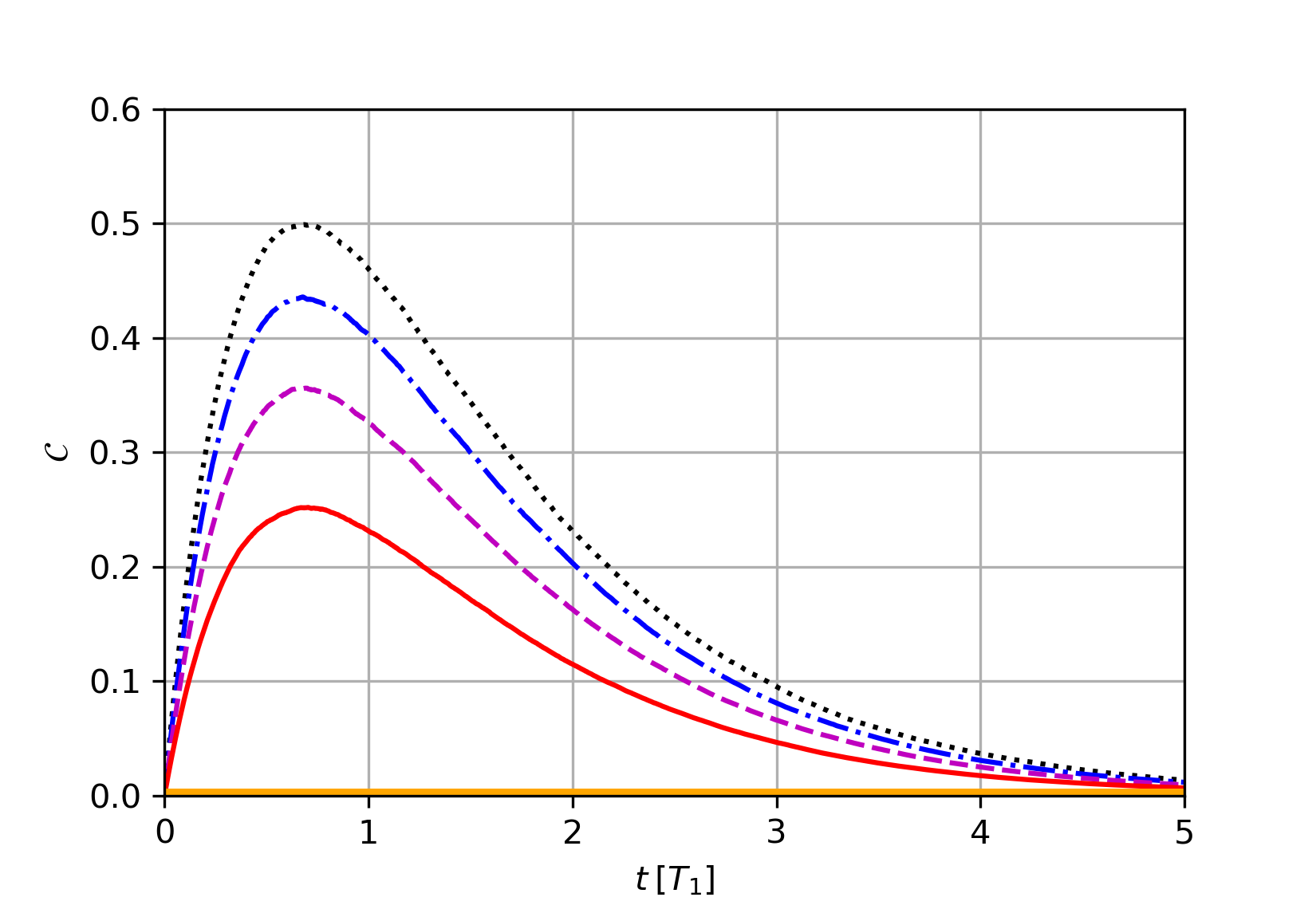} \\
    \begin{picture}(0,0)
    \put(48,139){$\vartheta - \theta = 90^\circ$}
    \put(48,127.5){\color{blue} $\vartheta - \theta = 60^\circ$}
    \put(48,116){\color{magenta} $\vartheta - \theta = 45^\circ$}
    \put(48,104.5){\color{red} $\vartheta - \theta = 30^\circ$}
    \put(48,93){\color{Orange} $\vartheta - \theta = 0^\circ$}
    \put(-90,158){(b)}
    \put(-90,310){(a)}
    \end{picture} \vspace{-15pt}
    \caption{We show the concurrence as a function of time, both for individual trajectories and averaged from an ensemble of them, as obtained from simulation of the double homodyne measurement depicted in Fig.~\ref{fig-splitterphase}(b). The initial state is $\ket{ee}$ for all trajectories. In (a) we show the average (dark blue) over individual stochastic trajectories for the settings $\theta = 0^\circ$ and $\vartheta = 90^\circ$. A pale blue envelope of $\pm$ one standard deviation surrounds the average. We see that some trajectories do much better than the average, with many reaching maximal entanglement $\mathcal{C} = 1$. The average concurrence from the comparable photodetection case \eqref{barC-jhe} (dotted red) is found to be in good agreement with the homodyne case. In (b) we plot the average concurrence from simulations, with different relationships between the phases $\theta$ and $\vartheta$; this demonstrates how the entanglement yield is affected by changing the relative phases of the two homodyne measurements. The optimal choice (dotted black, or the top panel) makes it impossible for inferences about the photon source to be drawn from the measurement record (the which--path information is erased), while the least--optimal choice (orange) maximizes the amount of information available about the photon source, and destroys the possibility of entanglement genesis entirely.}
    \label{fig-homodyne-eeGen}
\end{figure}

We proceed to one of our main results, namely that ideal homodyne detection, described by \eqref{hom-stateup}, can generate two--qubit entanglement given the initially--separable two--qubit state $\ket{ee}$. Unlike with the photodetection case however, the ability of a pair of quadrature measurements to erase information about which qubit is emitting a particular signal depends on the phases $\theta$ and $\vartheta$. This is apparent if we consider a probability density in terms of the quadrature readouts $X_3$ and $X_4$
\be \label{tQ-probdist} \begin{split}
\wp &= |\ip{X_3 X_4}{\psi_{3,4}}|^2 \\ & \propto e^{-X_3^2-X_4^2} \left( X_3^2 + X_4^2 \pm 2 X_3 X_4 \cos(\theta-\vartheta) \right).
\end{split} \ee
The state $\ket{\psi_{3,4}} = \tfrac{1}{\sqrt{2}}e^{i\theta} \ket{1_3 0_4} \pm \tfrac{1}{\sqrt{2}}e^{i\vartheta} \ket{0_3 1_4}$ represents the optical state in modes 3 and 4, where $+$ denotes the case where a single photon entered the beamsplitter from port 1 (with certainty), and $-$ corresponds to the case where it entered from port 2 (with certainty). 
When the $\pm$ distributions do not overlap completely, it is possible to make some inference about the which--qubit origin of different contributions to the signal based on the device readouts. This is however impossible when $\theta$ and $\vartheta$ are $90^\circ$ out of phase, such that $\cos(\theta-\vartheta)=0$ and the distributions overlap completely. 
Thus, we can erase the which--path information in a homodyne measurement by choosing the quadratures we measure at each output to be $90^\circ$ apart. 
This general measurement strategy on a pair of qubits has previously been shown to be optimal for entanglement preservation among diffusive measurement strategies, by completely different methods \cite{Viviescas2010}. 
It is possible to compute the concurrence after one timestep analytically, and we find that the concurrence of $\mathcal{M}_{34}\ket{ee}$ goes like $\mathcal{C} \propto |e^{2i\vartheta} - e^{2i\theta}|$; this shows that some entanglement is generated deterministically from $\ket{ee}$ in the first timestep (there is no dependence on $X_3$ and $X_4$ in this expression), and that the amount of concurrence generated in this manner is maximized by choosing $|\theta - \vartheta| = 90^\circ$ and thereby insuring the measurement erases which--path information. 

\par This optimal choice of quadratures corresponds to making an Einstein--Podolsky--Rosen (EPR) measurement \cite{EPR1935} on the optical modes, e.g.~we measure $\hat{X}_3|_{\theta = 0} = (\hat{X}_1+\hat{X}_2)/\sqrt{2}$, and $\hat{X}_4|_{\vartheta = 90^\circ} = (\hat{a}_4^\dag e^{i\vartheta} + \hat{a}_4 e^{-i\vartheta})/\sqrt{2}|_{\vartheta = 90^\circ}
$. 
Following this line of reasoning we can see that the entangling effect of the measurement can be understood as an entanglement swapping operation, where we begin with entanglement between each qubit and its optical mode, and swap it to appear between the two qubits instead. This interpretation has been applied to such homodyne measurements before \cite{Takeda2015}, and circuit--QED implementations moving towards such capabilities in the microwave regime have been proposed and realized \cite{Flurin2012, Silveri2016}.
Note that a heterodyne measurement, i.e.~one where we monitor \emph{both} quadratures at each output, rather than only one quadrature at each output, does not allow us to erase our which--path information; the optical states prepared by such a heterodyne measurement are separable, and do not perform any entanglement swapping or generate any subsequent two--qubit entanglement. The same argument can be applied to the non--entangling case of the homodyne measurement with $\theta = \vartheta$. See the supporting manuscript for details \cite{LongFlor2019}.
Similar principles apply for other measurements schemes. 
For example, dispersive monitoring is the most common alternative in circuit--QED to the fluorescence--based measurements we have described here; 
although dispersive monitoring involves a different interaction between the fields and qubits, and leads to different two--qubit dynamics, the need to erase which--path information in the device geometry we consider leads to effectively the same downstream entangling measurement \cite{Silveri2016} as we have described above.

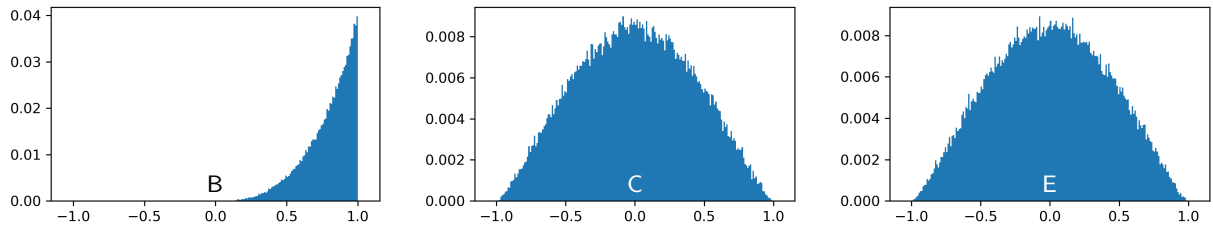
\begin{figure*}
\begin{tabular}{ccc}
\includegraphics[width =.3\textwidth]{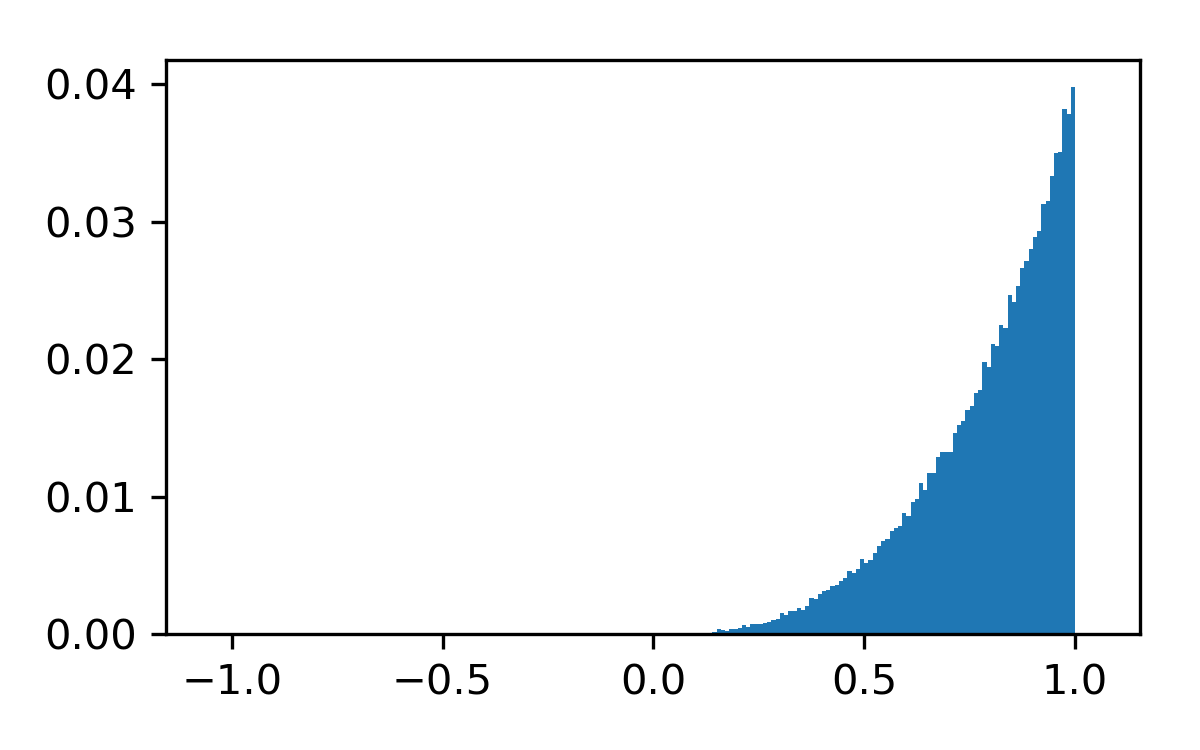} &
\includegraphics[width =.3\textwidth]{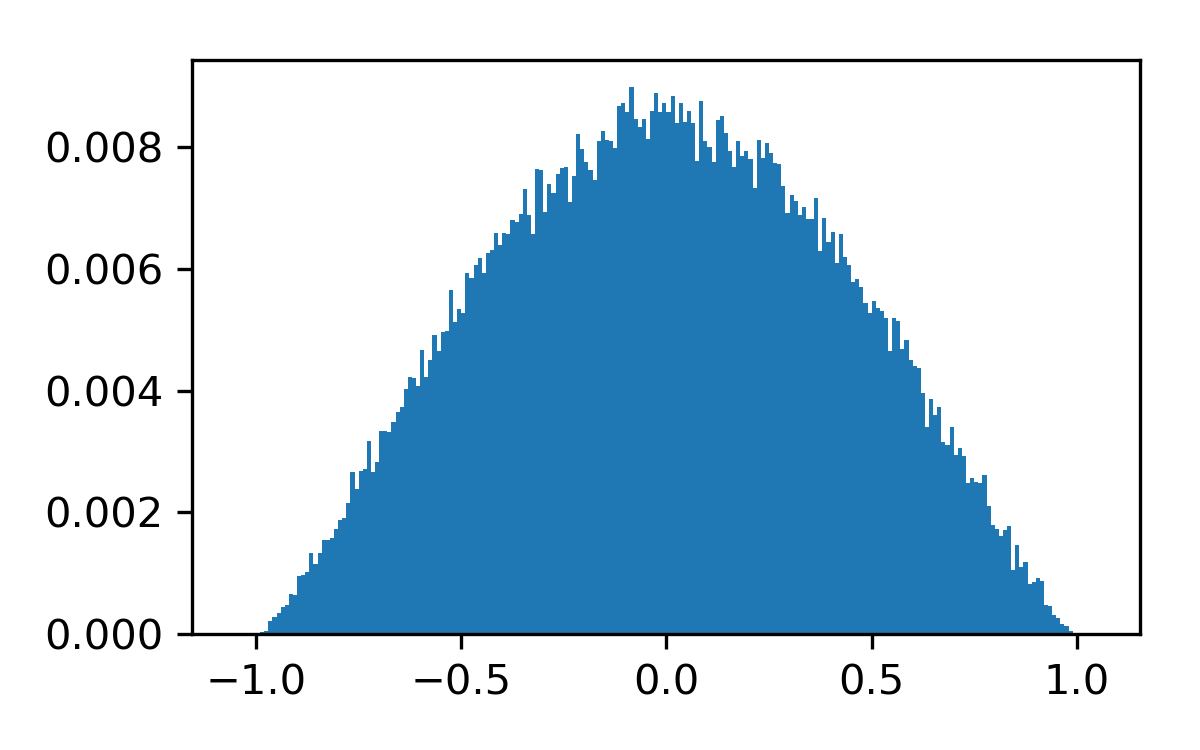} &
\includegraphics[width =.3\textwidth]{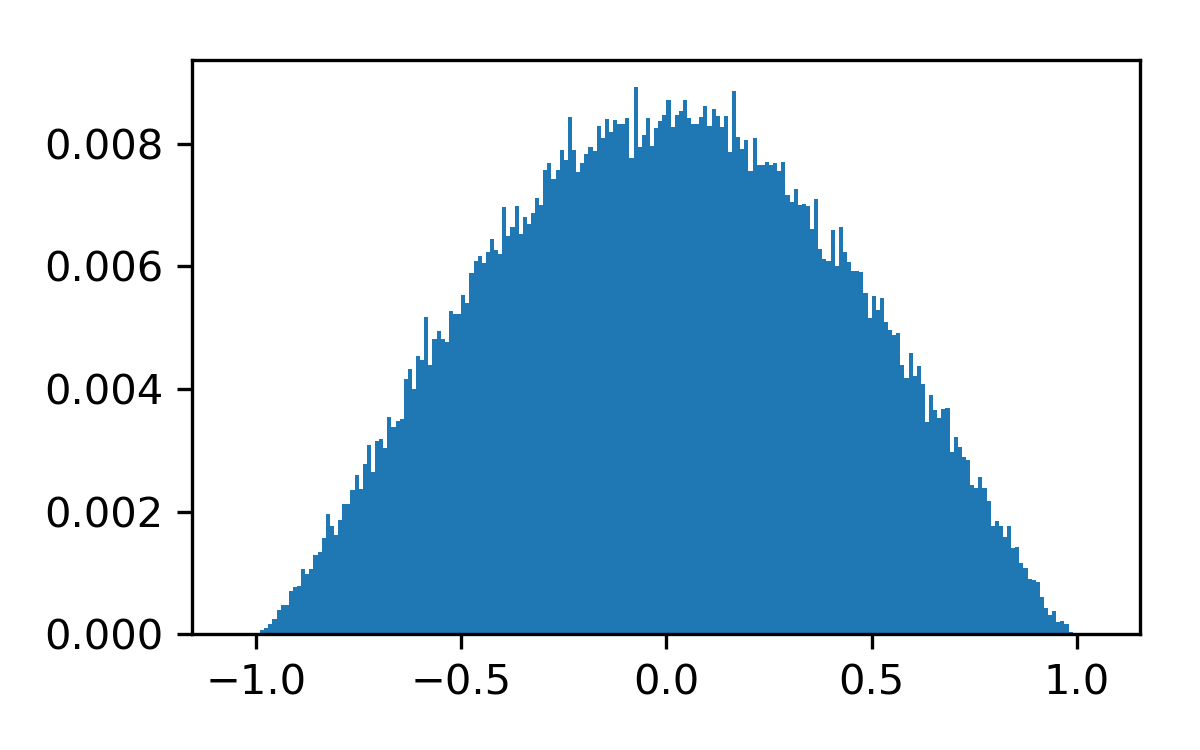}
\end{tabular} \vspace{-10pt} 
\begin{picture}(0,0)
\put(-392,-26){$\mathsf{B}$}
\put(-233,-26){\color{white}$\mathsf{C}$}
\put(-76,-26){\color{white}$\mathsf{E}$}
\end{picture}
\caption{We show histograms of the amplitudes $\mathsf{B}$, $\mathsf{C}$, and $\mathsf{E}$ defined in \eqref{BellSupState}, as obtained from simulations initialized at $\ket{ee}$, with $\theta = 0$ and $\vartheta = 90^\circ$. The distributions represent the states that occur in the first timestep of trajectory evolution for which successful trajectories exceed $\mathcal{C} = 0.999$; 100,000 such states are used to generate the histograms shown. All $y$--axes show normalized counts. We see symmetric distributions of $\mathsf{C}$ and $\mathsf{E}$, such that the entire range of $\mathsf{C}$, $\mathsf{E}$, and $\mathsf{B} >0$ is explored, and single most--likely state is denoted by $\mathsf{B} = 1$.}
\label{fig-states}
\end{figure*}

\par These points can be confirmed and extended by simulating SQTs under the continuous homodyne measurement dynamics. 
The concurrence of simulated individual trajectories and ensembles are shown in Fig.~\ref{fig-homodyne-eeGen}. We see that diffusive trajectories originating from $\ket{ee}$ do indeed gain concurrence deterministically early in the simulation, before diffusing apart based on different measurement records. 
For an optimal choice of quadratures, some trajectories reach the maximum $\mathcal{C} = 1$ in their evolution, with the peak of the average concurrence occuring at $\mathcal{C} = \tfrac{1}{2}$ and $t = T_1 \text{ln}(2)$. 
Thus this measurement can generate enough entanglement to be operationally useful, allowing us to probabilistically prepare maximally--entangled states by post--selecting on the $\mathcal{C} \approx 1$ trajectories. 
Tuning the measurement quadratures away from the optimal $|\theta - \vartheta| = 90^\circ$ preserves the shape of the curve \eqref{barC-jhe} for SQTs originating from $\ket{ee}$, but attenuates it, to the point where no two--qubit entanglement whatsoever is generated for the worst case $\theta = \vartheta$.

\par Contrary to the setups generating entanglement from a dispersive parity measurement or from photon counting, the entangled states generated by the process under study are not necessarily one of the four Bell states in the computational basis. At the times our diffusive quantum trajectories reach maximal concurrence, evolving from an initial state $\ket{ee}$ with $\theta = 0$ and $\vartheta = 90^\circ$, the two--qubit state can be in a superposition $\ket{\psi} = $
\be \label{BellSupState}
\tfrac{\mathsf{B}}{\sqrt{2}}(\ket{ee}-\ket{gg})+\tfrac{\mathsf{C}}{\sqrt{2}}(\ket{eg}+\ket{ge}) + i\tfrac{\mathsf{E}}{\sqrt{2}}(\ket{eg}-\ket{ge})
\ee
of several Bell states, which has $\mathsf{B}, \mathsf{C}, \mathsf{E} \in \mathbb{R}$ with $\mathsf{B} > 0$, and is still maximally entangled. The distribution of these maximally entangled states, at the first time--step in which successful trajectories reach $\mathcal{C} > 0.999$, is illustrated in Fig.~\ref{fig-states}. We emphasize however that the knowledge of the readout and qubit spontaneous emission rate is enough to perfectly track the two--qubit state and its concurrence, making the post--selection realistic. In this sense, this homomdyne entangling measurement is heralded. 
Any state of the type \eqref{BellSupState} can be converted to one of the standard Bell states (such as $\ket{\Phi^-}$, for which $\theta = 0$ and $\vartheta = 90^\circ$ then give the optimal measurement strategy for preserving concurrence \cite{LongFlor2019, Viviescas2010}) using a unitary operation on a single qubit.

\par Let us now compare these diffusive dynamics with the better--known photodetection case \cite{Mascarenhas2010, Mascarenhas2011, Carvalho2011, Santos2012}.
Photodetection events can project separable two--qubit states onto Bell states when a detector clicks, as understood from prior theory and experiments \cite{Cabrillo1998, Plenio1999, Moehring2007, Hofmann2012, Santos2012, Slodika2013, Hanson2013Heralded, Ohm2017, Araneda2018}. In this case, the 50/50 beamsplitter shown in Fig.~\ref{fig-splitterphase}(a) alone is sufficient to erase the which--path information of an emitted photon, such that e.g.~$\ket{ee}$ is projected onto a Bell state $\tfrac{1}{\sqrt{2}}\ket{eg} \pm \tfrac{1}{\sqrt{2}}\ket{ge}$ when a click is registered at output 3($+$) or 4($-$). 
Eventually a second photon will emerge in this scenario, and destroy the entangled state, with the qubits jumping to $\ket{gg}$. We cannot know from which qubit the second photon came any more than the first, but interference effects present in our model dictate that the sequential pair of clicks must occur on the same detector in any given realization.
The average evolution of the concurrence \cite{wooters1998} as a function of time can be determined analytically in this case, using relatively standard methods \cite{Mascarenhas2010, Mascarenhas2011, Santos2012}. 
If each qubit and its field mode evolves according to $\sqrt{e^{-\gamma t}} \ket{e0} + \sqrt{1-e^{-\gamma t}} \ket{g1}$, then a pair evolves according to 
\be \begin{split} \label{barC-jhe-derivation}
 \ket{\psi} = & e^{-\gamma t} \ket{ee}\ket{00} + \left( 1 - e^{-\gamma t}\right) \ket{gg}\ket{11} \\ 
  & + \sqrt{ e^{-\gamma t}(1-e^{-\gamma t})} \left[\ket{eg} \ket{01} + \ket{ge}\ket{10} \right],
\end{split} \ee
before the beamsplitter. Photodetections which project into the subspace of the term on the second line (after the beamsplitter) are those which take the two--qubit concurrence $\mathcal{C}$ \cite{wooters1998} from $0$ to $1$, such that the average of the concurrence over all jump trajectory realizations simply follows the probability to be in that subspace, i.e.
\be \label{barC-jhe}
\bar{\mathcal{C}}(t) = 2 e^{-\gamma t}(1-e^{-\gamma t}).
\ee
Further details and simulations of photodetection scenarios can be found in \cite{LongFlor2019}.
The same curve can be derived from the homodyne case, using eq.~(9) of \cite{Viviescas2010}, and the connection to the phodoetection case has been noted by \cite{Mascarenhas2011} (see the discussion around their eq.~(6)). For the initial state $\ket{ee}$, eq.~(9) of \cite{Viviescas2010} becomes 
\be 
\dot{\bar{\mathcal{C}}} = - \gamma \bar{\mathcal{C}} + 2 \gamma e^{-2\gamma t},
\ee
to which \eqref{barC-jhe} is a solution. We draw a number of conclusions from this: 1) the best homodyne scheme can be proven to have the same average entanglement yield as the comparable photodetection setup, consistent with the numerics shown in Fig.~\ref{fig-homodyne-eeGen} and in \cite{LongFlor2019}. 2) Such a result is intuitive once both the photodetection scheme \cite{Santos2012} and the homodyne scheme \cite{Takeda2015} are recognized as implementing entanglement swaps. 3) Despite the exact correspondence of the two approaches \emph{on average}, the trajectories and two--qubit states generated in individual realizations are of a completely different character. 

\par We have re--proposed \cite{Viviescas2010} a method of jointly monitoring the fluorescence of two qubits in spatially--separated cavities, involving a homodyne measurement of two fluorescence signals after they are mixed on a beamsplitter. 
Such measurements on single qubits have recently been realized in the laboratory \cite{PCI-2013, Jordan2015flor, Campagne-Ibarcq2016, PCI-2016-2, Naghiloo2016flor, Mahdi2016, Tan2017, Ficheux2018}. 
The degree to which the joint homodyne measurement can distinguish which qubit makes particular contributions to the monitored signals is based on the phase relationship between the quadratures monitored at each output, and is thus completely tunable. 
From previous works considering circuit--QED implementations of continuous measurement--based entanglement schemes, it is clear that making certain states indistinguishable with respect to the measurement used is a key to entanglement generation. 
In parity measurements \cite{Martin_2017}, utilizing either two qubits in a single cavity \cite{Riste2013}, or a pair of cavities probed dispersively in series \cite{Roch2014, martin2015remote, Motzoi2015, Chantasri2016}, this means that a certain parity outcome superposes a pair of two--qubit states which it cannot distinguish. When the cavities/emitters are in parallel, as in Fig.~\ref{fig-splitterphase} or similar \cite{Cabrillo1998, Plenio1999, Moehring2007, Hofmann2012, Slodika2013, Hanson2013Heralded, Silveri2016, Ohm2017}, this principle is manifest in erasing the source of any given signal, as we have discussed above. 
The entanglement generation mechanism in schemes with a parallel device geometry, particularly when monitoring spontaneous emission (again, see Fig.~\ref{fig-splitterphase}), contain challenging quantum elements; the qubits emit photons which go to a detector, and the emitters become entangled without ever having directly interacted (i.e.~they interact only through measurement backaction, which itself acts through entanglement between the qubits of interest and mediating field mode, carrying information to an observer).
Yet, by carefully choosing the kind of complete information collected (i.e.~engineering their ignorance in a particular way), an observer can establish useful coherent quantum (anti--)correlations between the emitters. 

While remote entanglement by photodetection has become standard practice at optical wavelengths, experiments in the microwave regime generally lack good photodetection capabilities; our results demonstrate the possibility of generating qubit entanglement in these regimes based on diffusive quantum trajectories, using natural extensions of, or applications of, existing amplification and measurement techniques \cite{Flurin2012, Silveri2016}. 
Advances in continuous measurements imply the possibility of new forms of feedback control; in this case there are photodetection--based works \cite{BarrettKok, Mascarenhas2010, Mascarenhas2010-1}, and more--recently homodyne--based work \cite{Leigh2019}, in this direction, which aim to increase the entanglement yield or extend the lifetime of entangled states using such methods. 

\par \emph{Acknowledgements:} We acknowledge helpful conversations with Benjamin Huard and Alexander N. Korotkov. We are grateful to Marcelo F.~Santos for pointing out a number of important references, including \cite{Viviescas2010, Mascarenhas2011}, after we posted v1 of our pre--print. PL, CE, SKM, and ANJ acknowledge funding from NSF grant no.~DMR-1809343, and US Army Research Office grant no.~W911NF-18-10178. PL acknowledges additional support from the US Department of Education grant No.~GR506598 as a GAANN fellow, and thanks the Quantum Information Machines school at \'{E}cole de Physique des Houches for their hospitality during part of this manuscript's preparation. XFQ and JHE acknowledge support from NSF grants PHY-1505189 and PHY-1539859.

\bibliography{refs}

\begin{thebibliography}{69}%
\makeatletter
\providecommand \@ifxundefined [1]{%
 \@ifx{#1\undefined}
}%
\providecommand \@ifnum [1]{%
 \ifnum #1\expandafter \@firstoftwo
 \else \expandafter \@secondoftwo
 \fi
}%
\providecommand \@ifx [1]{%
 \ifx #1\expandafter \@firstoftwo
 \else \expandafter \@secondoftwo
 \fi
}%
\providecommand \natexlab [1]{#1}%
\providecommand \enquote  [1]{``#1''}%
\providecommand \bibnamefont  [1]{#1}%
\providecommand \bibfnamefont [1]{#1}%
\providecommand \citenamefont [1]{#1}%
\providecommand \href@noop [0]{\@secondoftwo}%
\providecommand \href [0]{\begingroup \@sanitize@url \@href}%
\providecommand \@href[1]{\@@startlink{#1}\@@href}%
\providecommand \@@href[1]{\endgroup#1\@@endlink}%
\providecommand \@sanitize@url [0]{\catcode `\\12\catcode `\$12\catcode
  `\&12\catcode `\#12\catcode `\^12\catcode `\_12\catcode `\%12\relax}%
\providecommand \@@startlink[1]{}%
\providecommand \@@endlink[0]{}%
\providecommand \url  [0]{\begingroup\@sanitize@url \@url }%
\providecommand \@url [1]{\endgroup\@href {#1}{\urlprefix }}%
\providecommand \urlprefix  [0]{URL }%
\providecommand \Eprint [0]{\href }%
\providecommand \doibase [0]{http://dx.doi.org/}%
\providecommand \selectlanguage [0]{\@gobble}%
\providecommand \bibinfo  [0]{\@secondoftwo}%
\providecommand \bibfield  [0]{\@secondoftwo}%
\providecommand \translation [1]{[#1]}%
\providecommand \BibitemOpen [0]{}%
\providecommand \bibitemStop [0]{}%
\providecommand \bibitemNoStop [0]{.\EOS\space}%
\providecommand \EOS [0]{\spacefactor3000\relax}%
\providecommand \BibitemShut  [1]{\csname bibitem#1\endcsname}%
\let\auto@bib@innerbib\@empty
\bibitem [{\citenamefont {Carmichael}(1993)}]{BookCarmichael}%
  \BibitemOpen
  \bibfield  {author} {\bibinfo {author} {\bibfnamefont {H.~J.}\ \bibnamefont
  {Carmichael}},\ }\href@noop {} {\emph {\bibinfo {title} {An Open Systems
  Approach to Quantum Optics}}}\ (\bibinfo  {publisher} {Springer, Berlin},\
  \bibinfo {year} {1993})\BibitemShut {NoStop}%
\bibitem [{\citenamefont {Percival}(1998)}]{BookPercival}%
  \BibitemOpen
  \bibfield  {author} {\bibinfo {author} {\bibfnamefont {I.~C.}\ \bibnamefont
  {Percival}},\ }\href {https://books.google.com/books?id=AlXSmZTHxtwC} {\emph
  {\bibinfo {title} {Quantum State Diffusion}}}\ (\bibinfo  {publisher}
  {Cambridge University Press},\ \bibinfo {year} {1998})\BibitemShut {NoStop}%
\bibitem [{\citenamefont {Gardiner}\ and\ \citenamefont
  {Zoller}(2004)}]{BookGardiner}%
  \BibitemOpen
  \bibfield  {author} {\bibinfo {author} {\bibfnamefont {C.~W.}\ \bibnamefont
  {Gardiner}}\ and\ \bibinfo {author} {\bibfnamefont {P.}~\bibnamefont
  {Zoller}},\ }\href@noop {} {\emph {\bibinfo {title} {{Quantum Noise: A
  Handbook of Markovian and Non-Markovian Quantum Stochastic Methods with
  Applications to Quantum Optics}}}}\ (\bibinfo  {publisher} {Springer},\
  \bibinfo {year} {2004})\BibitemShut {NoStop}%
\bibitem [{\citenamefont {Wiseman}\ and\ \citenamefont
  {Milburn}(2010)}]{BookWiseman}%
  \BibitemOpen
  \bibfield  {author} {\bibinfo {author} {\bibfnamefont {H.~M.}\ \bibnamefont
  {Wiseman}}\ and\ \bibinfo {author} {\bibfnamefont {G.~J.}\ \bibnamefont
  {Milburn}},\ }\href@noop {} {\emph {\bibinfo {title} {Quantum measurement and
  control}}}\ (\bibinfo  {publisher} {Cambridge University Press},\ \bibinfo
  {year} {2010})\BibitemShut {NoStop}%
\bibitem [{\citenamefont {Barchielli}\ and\ \citenamefont
  {Gregoratti}(2009)}]{BookBarchielli}%
  \BibitemOpen
  \bibfield  {author} {\bibinfo {author} {\bibfnamefont {A.}~\bibnamefont
  {Barchielli}}\ and\ \bibinfo {author} {\bibfnamefont {M.}~\bibnamefont
  {Gregoratti}},\ }\href@noop {} {\emph {\bibinfo {title} {Quantum trajectories
  and measurements in continuous time}}}\ (\bibinfo  {publisher}
  {Springer-Verlag Berlin Heidelberg},\ \bibinfo {year} {2009})\BibitemShut
  {NoStop}%
\bibitem [{\citenamefont {Wiseman}(1996)}]{Wiseman1996Review}%
  \BibitemOpen
  \bibfield  {author} {\bibinfo {author} {\bibfnamefont {H.~M.}\ \bibnamefont
  {Wiseman}},\ }\href
  {https://iopscience.iop.org/article/10.1088/1355-5111/8/1/015} {\bibfield
  {journal} {\bibinfo  {journal} {Quantum and Semiclassical Optics: Journal of
  the European Optical Society Part B}\ }\textbf {\bibinfo {volume} {8}},\
  \bibinfo {pages} {205} (\bibinfo {year} {1996})}\BibitemShut {NoStop}%
\bibitem [{\citenamefont {Brun}(2002)}]{Brun2001Teach}%
  \BibitemOpen
  \bibfield  {author} {\bibinfo {author} {\bibfnamefont {T.~A.}\ \bibnamefont
  {Brun}},\ }\href {\doibase 10.1119/1.1475328} {\bibfield  {journal} {\bibinfo
   {journal} {{American Journal of Physics}}\ }\textbf {\bibinfo {volume}
  {70}},\ \bibinfo {pages} {719} (\bibinfo {year} {{2002}})}\BibitemShut
  {NoStop}%
\bibitem [{\citenamefont {Jacobs}\ and\ \citenamefont
  {Steck}(2006)}]{Jacobs2006}%
  \BibitemOpen
  \bibfield  {author} {\bibinfo {author} {\bibfnamefont {K.}~\bibnamefont
  {Jacobs}}\ and\ \bibinfo {author} {\bibfnamefont {D.~A.}\ \bibnamefont
  {Steck}},\ }\href {\doibase 10.1080/00107510601101934} {\bibfield  {journal}
  {\bibinfo  {journal} {Contemporary Physics}\ }\textbf {\bibinfo {volume}
  {47}},\ \bibinfo {pages} {279} (\bibinfo {year} {2006})}\BibitemShut
  {NoStop}%
\bibitem [{\citenamefont {Chantasri}\ \emph {et~al.}(2013)\citenamefont
  {Chantasri}, \citenamefont {Dressel},\ and\ \citenamefont
  {Jordan}}]{Chantasri2013}%
  \BibitemOpen
  \bibfield  {author} {\bibinfo {author} {\bibfnamefont {A.}~\bibnamefont
  {Chantasri}}, \bibinfo {author} {\bibfnamefont {J.}~\bibnamefont {Dressel}},
  \ and\ \bibinfo {author} {\bibfnamefont {A.~N.}\ \bibnamefont {Jordan}},\
  }\href {\doibase 10.1103/PhysRevA.88.042110} {\bibfield  {journal} {\bibinfo
  {journal} {Phys. Rev. A}\ }\textbf {\bibinfo {volume} {88}},\ \bibinfo
  {pages} {042110} (\bibinfo {year} {2013})}\BibitemShut {NoStop}%
\bibitem [{\citenamefont {Chantasri}\ and\ \citenamefont
  {Jordan}(2015)}]{Chantasri2015}%
  \BibitemOpen
  \bibfield  {author} {\bibinfo {author} {\bibfnamefont {A.}~\bibnamefont
  {Chantasri}}\ and\ \bibinfo {author} {\bibfnamefont {A.~N.}\ \bibnamefont
  {Jordan}},\ }\href {\doibase 10.1103/PhysRevA.92.032125} {\bibfield
  {journal} {\bibinfo  {journal} {Phys. Rev. A}\ }\textbf {\bibinfo {volume}
  {92}},\ \bibinfo {pages} {032125} (\bibinfo {year} {2015})}\BibitemShut
  {NoStop}%
\bibitem [{\citenamefont {Korotkov}(2016)}]{Korotkov2016}%
  \BibitemOpen
  \bibfield  {author} {\bibinfo {author} {\bibfnamefont {A.~N.}\ \bibnamefont
  {Korotkov}},\ }\href {\doibase 10.1103/PhysRevA.94.042326} {\bibfield
  {journal} {\bibinfo  {journal} {Phys. Rev. A}\ }\textbf {\bibinfo {volume}
  {94}},\ \bibinfo {pages} {042326} (\bibinfo {year} {2016})}\BibitemShut
  {NoStop}%
\bibitem [{\citenamefont {Gambetta}\ \emph {et~al.}(2008)\citenamefont
  {Gambetta}, \citenamefont {Blais}, \citenamefont {Boissonneault},
  \citenamefont {Houck}, \citenamefont {Schuster},\ and\ \citenamefont
  {Girvin}}]{Gambetta2008}%
  \BibitemOpen
  \bibfield  {author} {\bibinfo {author} {\bibfnamefont {J.}~\bibnamefont
  {Gambetta}}, \bibinfo {author} {\bibfnamefont {A.}~\bibnamefont {Blais}},
  \bibinfo {author} {\bibfnamefont {M.}~\bibnamefont {Boissonneault}}, \bibinfo
  {author} {\bibfnamefont {A.~A.}\ \bibnamefont {Houck}}, \bibinfo {author}
  {\bibfnamefont {D.~I.}\ \bibnamefont {Schuster}}, \ and\ \bibinfo {author}
  {\bibfnamefont {S.~M.}\ \bibnamefont {Girvin}},\ }\href {\doibase
  10.1103/PhysRevA.77.012112} {\bibfield  {journal} {\bibinfo  {journal} {Phys.
  Rev. A}\ }\textbf {\bibinfo {volume} {77}},\ \bibinfo {pages} {012112}
  (\bibinfo {year} {2008})}\BibitemShut {NoStop}%
\bibitem [{\citenamefont {Murch}\ \emph {et~al.}(2013)\citenamefont {Murch},
  \citenamefont {Weber}, \citenamefont {Macklin},\ and\ \citenamefont
  {Siddiqi}}]{Murch2013}%
  \BibitemOpen
  \bibfield  {author} {\bibinfo {author} {\bibfnamefont {K.~W.}\ \bibnamefont
  {Murch}}, \bibinfo {author} {\bibfnamefont {S.~J.}\ \bibnamefont {Weber}},
  \bibinfo {author} {\bibfnamefont {C.}~\bibnamefont {Macklin}}, \ and\
  \bibinfo {author} {\bibfnamefont {I.}~\bibnamefont {Siddiqi}},\ }\href
  {https://www.nature.com/articles/nature12539} {\bibfield  {journal} {\bibinfo
   {journal} {Nature}\ }\textbf {\bibinfo {volume} {502}},\ \bibinfo {pages}
  {211} (\bibinfo {year} {2013})}\BibitemShut {NoStop}%
\bibitem [{\citenamefont {Weber}\ \emph {et~al.}(2014)\citenamefont {Weber},
  \citenamefont {Chantasri}, \citenamefont {Dressel}, \citenamefont {Jordan},
  \citenamefont {Murch},\ and\ \citenamefont {Siddiqi}}]{Weber2014}%
  \BibitemOpen
  \bibfield  {author} {\bibinfo {author} {\bibfnamefont {S.~J.}\ \bibnamefont
  {Weber}}, \bibinfo {author} {\bibfnamefont {A.}~\bibnamefont {Chantasri}},
  \bibinfo {author} {\bibfnamefont {J.}~\bibnamefont {Dressel}}, \bibinfo
  {author} {\bibfnamefont {A.~N.}\ \bibnamefont {Jordan}}, \bibinfo {author}
  {\bibfnamefont {K.~W.}\ \bibnamefont {Murch}}, \ and\ \bibinfo {author}
  {\bibfnamefont {I.}~\bibnamefont {Siddiqi}},\ }\href {\doibase
  10.1038/nature13559} {\bibfield  {journal} {\bibinfo  {journal} {Nature}\
  }\textbf {\bibinfo {volume} {511}},\ \bibinfo {pages} {570} (\bibinfo {year}
  {2014})}\BibitemShut {NoStop}%
\bibitem [{\citenamefont {Caves}(1982)}]{Caves1982}%
  \BibitemOpen
  \bibfield  {author} {\bibinfo {author} {\bibfnamefont {C.~M.}\ \bibnamefont
  {Caves}},\ }\href {\doibase 10.1103/PhysRevD.26.1817} {\bibfield  {journal}
  {\bibinfo  {journal} {Phys. Rev. D}\ }\textbf {\bibinfo {volume} {26}},\
  \bibinfo {pages} {1817} (\bibinfo {year} {1982})}\BibitemShut {NoStop}%
\bibitem [{\citenamefont {Clerk}\ \emph {et~al.}(2010)\citenamefont {Clerk},
  \citenamefont {Devoret}, \citenamefont {Girvin}, \citenamefont {Marquardt},\
  and\ \citenamefont {Schoelkopf}}]{Clerk2010Review}%
  \BibitemOpen
  \bibfield  {author} {\bibinfo {author} {\bibfnamefont {A.~A.}\ \bibnamefont
  {Clerk}}, \bibinfo {author} {\bibfnamefont {M.~H.}\ \bibnamefont {Devoret}},
  \bibinfo {author} {\bibfnamefont {S.~M.}\ \bibnamefont {Girvin}}, \bibinfo
  {author} {\bibfnamefont {F.}~\bibnamefont {Marquardt}}, \ and\ \bibinfo
  {author} {\bibfnamefont {R.~J.}\ \bibnamefont {Schoelkopf}},\ }\href
  {\doibase 10.1103/RevModPhys.82.1155} {\bibfield  {journal} {\bibinfo
  {journal} {Rev. Mod. Phys.}\ }\textbf {\bibinfo {volume} {82}},\ \bibinfo
  {pages} {1155} (\bibinfo {year} {2010})}\BibitemShut {NoStop}%
\bibitem [{\citenamefont {Bergeal}\ \emph {et~al.}(2010)\citenamefont
  {Bergeal}, \citenamefont {Schackert}, \citenamefont {Metcalfe}, \citenamefont
  {Vijay}, \citenamefont {Manucharyan}, \citenamefont {Frunzio}, \citenamefont
  {Prober}, \citenamefont {Schoelkopf}, \citenamefont {Girvin},\ and\
  \citenamefont {Devoret}}]{Bergeal2010}%
  \BibitemOpen
  \bibfield  {author} {\bibinfo {author} {\bibfnamefont {N.}~\bibnamefont
  {Bergeal}}, \bibinfo {author} {\bibfnamefont {F.}~\bibnamefont {Schackert}},
  \bibinfo {author} {\bibfnamefont {M.}~\bibnamefont {Metcalfe}}, \bibinfo
  {author} {\bibfnamefont {R.}~\bibnamefont {Vijay}}, \bibinfo {author}
  {\bibfnamefont {V.~E.}\ \bibnamefont {Manucharyan}}, \bibinfo {author}
  {\bibfnamefont {L.}~\bibnamefont {Frunzio}}, \bibinfo {author} {\bibfnamefont
  {D.~E.}\ \bibnamefont {Prober}}, \bibinfo {author} {\bibfnamefont {R.~J.}\
  \bibnamefont {Schoelkopf}}, \bibinfo {author} {\bibfnamefont {S.~M.}\
  \bibnamefont {Girvin}}, \ and\ \bibinfo {author} {\bibfnamefont {M.~H.}\
  \bibnamefont {Devoret}},\ }\href
  {https://www.nature.com/articles/nature09035} {\bibfield  {journal} {\bibinfo
   {journal} {Nature}\ }\textbf {\bibinfo {volume} {465}},\ \bibinfo {pages}
  {6} (\bibinfo {year} {2010})}\BibitemShut {NoStop}%
\bibitem [{\citenamefont {Roy}\ and\ \citenamefont
  {Devoret}(2016)}]{DevoretAmp}%
  \BibitemOpen
  \bibfield  {author} {\bibinfo {author} {\bibfnamefont {A.}~\bibnamefont
  {Roy}}\ and\ \bibinfo {author} {\bibfnamefont {M.}~\bibnamefont {Devoret}},\
  }\href
  {https://www.sciencedirect.com/science/article/pii/S1631070516300640?via%3Dihub}
  {\bibfield  {journal} {\bibinfo  {journal} {Comptes Rendus Physique}\
  }\textbf {\bibinfo {volume} {17}},\ \bibinfo {pages} {740 } (\bibinfo {year}
  {2016})}\BibitemShut {NoStop}%
\bibitem [{\citenamefont {Ruskov}\ and\ \citenamefont
  {Korotkov}(2003)}]{Ruskov2003-2}%
  \BibitemOpen
  \bibfield  {author} {\bibinfo {author} {\bibfnamefont {R.}~\bibnamefont
  {Ruskov}}\ and\ \bibinfo {author} {\bibfnamefont {A.~N.}\ \bibnamefont
  {Korotkov}},\ }\href {\doibase 10.1103/PhysRevB.67.241305} {\bibfield
  {journal} {\bibinfo  {journal} {Phys. Rev. B}\ }\textbf {\bibinfo {volume}
  {67}},\ \bibinfo {pages} {241305} (\bibinfo {year} {2003})}\BibitemShut
  {NoStop}%
\bibitem [{\citenamefont {Trauzettel}\ \emph {et~al.}(2006)\citenamefont
  {Trauzettel}, \citenamefont {Jordan}, \citenamefont {Beenakker},\ and\
  \citenamefont {B\"uttiker}}]{Trauzettel2006}%
  \BibitemOpen
  \bibfield  {author} {\bibinfo {author} {\bibfnamefont {B.}~\bibnamefont
  {Trauzettel}}, \bibinfo {author} {\bibfnamefont {A.~N.}\ \bibnamefont
  {Jordan}}, \bibinfo {author} {\bibfnamefont {C.~W.~J.}\ \bibnamefont
  {Beenakker}}, \ and\ \bibinfo {author} {\bibfnamefont {M.}~\bibnamefont
  {B\"uttiker}},\ }\href {\doibase 10.1103/PhysRevB.73.235331} {\bibfield
  {journal} {\bibinfo  {journal} {Phys. Rev. B}\ }\textbf {\bibinfo {volume}
  {73}},\ \bibinfo {pages} {235331} (\bibinfo {year} {2006})}\BibitemShut
  {NoStop}%
\bibitem [{\citenamefont {Williams}\ and\ \citenamefont
  {Jordan}(2008)}]{Williams2008}%
  \BibitemOpen
  \bibfield  {author} {\bibinfo {author} {\bibfnamefont {N.~S.}\ \bibnamefont
  {Williams}}\ and\ \bibinfo {author} {\bibfnamefont {A.~N.}\ \bibnamefont
  {Jordan}},\ }\href {\doibase 10.1103/PhysRevA.78.062322} {\bibfield
  {journal} {\bibinfo  {journal} {Phys. Rev. A}\ }\textbf {\bibinfo {volume}
  {78}},\ \bibinfo {pages} {062322} (\bibinfo {year} {2008})}\BibitemShut
  {NoStop}%
\bibitem [{\citenamefont {Rist\`{e}}\ \emph {et~al.}(2013)\citenamefont
  {Rist\`{e}}, \citenamefont {Dukalski}, \citenamefont {Watson}, \citenamefont
  {de~Lange}, \citenamefont {Tiggelman}, \citenamefont {Blanter}, \citenamefont
  {Lehnert}, \citenamefont {Schouten},\ and\ \citenamefont
  {DiCarlo}}]{Riste2013}%
  \BibitemOpen
  \bibfield  {author} {\bibinfo {author} {\bibfnamefont {D.}~\bibnamefont
  {Rist\`{e}}}, \bibinfo {author} {\bibfnamefont {M.}~\bibnamefont {Dukalski}},
  \bibinfo {author} {\bibfnamefont {C.~A.}\ \bibnamefont {Watson}}, \bibinfo
  {author} {\bibfnamefont {G.}~\bibnamefont {de~Lange}}, \bibinfo {author}
  {\bibfnamefont {M.~J.}\ \bibnamefont {Tiggelman}}, \bibinfo {author}
  {\bibfnamefont {Y.~M.}\ \bibnamefont {Blanter}}, \bibinfo {author}
  {\bibfnamefont {K.~W.}\ \bibnamefont {Lehnert}}, \bibinfo {author}
  {\bibfnamefont {R.~N.}\ \bibnamefont {Schouten}}, \ and\ \bibinfo {author}
  {\bibfnamefont {L.}~\bibnamefont {DiCarlo}},\ }\href {\doibase
  10.1038/nature12513} {\bibfield  {journal} {\bibinfo  {journal} {Nature}\
  }\textbf {\bibinfo {volume} {502}},\ \bibinfo {pages} {350} (\bibinfo {year}
  {2013})}\BibitemShut {NoStop}%
\bibitem [{\citenamefont {Roch}\ \emph {et~al.}(2014)\citenamefont {Roch},
  \citenamefont {Schwartz}, \citenamefont {Motzoi}, \citenamefont {Macklin},
  \citenamefont {Vijay}, \citenamefont {Eddins}, \citenamefont {Korotkov},
  \citenamefont {Whaley}, \citenamefont {Sarovar},\ and\ \citenamefont
  {Siddiqi}}]{Roch2014}%
  \BibitemOpen
  \bibfield  {author} {\bibinfo {author} {\bibfnamefont {N.}~\bibnamefont
  {Roch}}, \bibinfo {author} {\bibfnamefont {M.~E.}\ \bibnamefont {Schwartz}},
  \bibinfo {author} {\bibfnamefont {F.}~\bibnamefont {Motzoi}}, \bibinfo
  {author} {\bibfnamefont {C.}~\bibnamefont {Macklin}}, \bibinfo {author}
  {\bibfnamefont {R.}~\bibnamefont {Vijay}}, \bibinfo {author} {\bibfnamefont
  {A.~W.}\ \bibnamefont {Eddins}}, \bibinfo {author} {\bibfnamefont {A.~N.}\
  \bibnamefont {Korotkov}}, \bibinfo {author} {\bibfnamefont {K.~B.}\
  \bibnamefont {Whaley}}, \bibinfo {author} {\bibfnamefont {M.}~\bibnamefont
  {Sarovar}}, \ and\ \bibinfo {author} {\bibfnamefont {I.}~\bibnamefont
  {Siddiqi}},\ }\href {\doibase 10.1103/PhysRevLett.112.170501} {\bibfield
  {journal} {\bibinfo  {journal} {Phys. Rev. Lett.}\ }\textbf {\bibinfo
  {volume} {112}},\ \bibinfo {pages} {170501} (\bibinfo {year}
  {2014})}\BibitemShut {NoStop}%
\bibitem [{\citenamefont {Martin}\ \emph {et~al.}(2015)\citenamefont {Martin},
  \citenamefont {Motzoi}, \citenamefont {Li}, \citenamefont {Sarovar},\ and\
  \citenamefont {Whaley}}]{martin2015remote}%
  \BibitemOpen
  \bibfield  {author} {\bibinfo {author} {\bibfnamefont {L.}~\bibnamefont
  {Martin}}, \bibinfo {author} {\bibfnamefont {F.}~\bibnamefont {Motzoi}},
  \bibinfo {author} {\bibfnamefont {H.}~\bibnamefont {Li}}, \bibinfo {author}
  {\bibfnamefont {M.}~\bibnamefont {Sarovar}}, \ and\ \bibinfo {author}
  {\bibfnamefont {K.~B.}\ \bibnamefont {Whaley}},\ }\href {\doibase
  10.1103/PhysRevA.92.062321} {\bibfield  {journal} {\bibinfo  {journal} {Phys.
  Rev. A}\ }\textbf {\bibinfo {volume} {92}},\ \bibinfo {pages} {062321}
  (\bibinfo {year} {2015})}\BibitemShut {NoStop}%
\bibitem [{\citenamefont {Motzoi}\ \emph {et~al.}(2015)\citenamefont {Motzoi},
  \citenamefont {Whaley},\ and\ \citenamefont {Sarovar}}]{Motzoi2015}%
  \BibitemOpen
  \bibfield  {author} {\bibinfo {author} {\bibfnamefont {F.}~\bibnamefont
  {Motzoi}}, \bibinfo {author} {\bibfnamefont {K.~B.}\ \bibnamefont {Whaley}},
  \ and\ \bibinfo {author} {\bibfnamefont {M.}~\bibnamefont {Sarovar}},\ }\href
  {\doibase 10.1103/PhysRevA.92.032308} {\bibfield  {journal} {\bibinfo
  {journal} {Phys. Rev. A}\ }\textbf {\bibinfo {volume} {92}},\ \bibinfo
  {pages} {032308} (\bibinfo {year} {2015})}\BibitemShut {NoStop}%
\bibitem [{\citenamefont {Silveri}\ \emph {et~al.}(2016)\citenamefont
  {Silveri}, \citenamefont {Zalys-Geller}, \citenamefont {Hatridge},
  \citenamefont {Leghtas}, \citenamefont {Devoret},\ and\ \citenamefont
  {Girvin}}]{Silveri2016}%
  \BibitemOpen
  \bibfield  {author} {\bibinfo {author} {\bibfnamefont {M.}~\bibnamefont
  {Silveri}}, \bibinfo {author} {\bibfnamefont {E.}~\bibnamefont
  {Zalys-Geller}}, \bibinfo {author} {\bibfnamefont {M.}~\bibnamefont
  {Hatridge}}, \bibinfo {author} {\bibfnamefont {Z.}~\bibnamefont {Leghtas}},
  \bibinfo {author} {\bibfnamefont {M.~H.}\ \bibnamefont {Devoret}}, \ and\
  \bibinfo {author} {\bibfnamefont {S.~M.}\ \bibnamefont {Girvin}},\ }\href
  {\doibase 10.1103/PhysRevA.93.062310} {\bibfield  {journal} {\bibinfo
  {journal} {Phys. Rev. A}\ }\textbf {\bibinfo {volume} {93}},\ \bibinfo
  {pages} {062310} (\bibinfo {year} {2016})}\BibitemShut {NoStop}%
\bibitem [{\citenamefont {Chantasri}\ \emph {et~al.}(2016)\citenamefont
  {Chantasri}, \citenamefont {Kimchi-Schwartz}, \citenamefont {Roch},
  \citenamefont {Siddiqi},\ and\ \citenamefont {Jordan}}]{Chantasri2016}%
  \BibitemOpen
  \bibfield  {author} {\bibinfo {author} {\bibfnamefont {A.}~\bibnamefont
  {Chantasri}}, \bibinfo {author} {\bibfnamefont {M.~E.}\ \bibnamefont
  {Kimchi-Schwartz}}, \bibinfo {author} {\bibfnamefont {N.}~\bibnamefont
  {Roch}}, \bibinfo {author} {\bibfnamefont {I.}~\bibnamefont {Siddiqi}}, \
  and\ \bibinfo {author} {\bibfnamefont {A.~N.}\ \bibnamefont {Jordan}},\
  }\href {\doibase 10.1103/PhysRevX.6.041052} {\bibfield  {journal} {\bibinfo
  {journal} {Phys. Rev. X}\ }\textbf {\bibinfo {volume} {6}},\ \bibinfo {pages}
  {041052} (\bibinfo {year} {2016})}\BibitemShut {NoStop}%
\bibitem [{\citenamefont {Martin}\ \emph {et~al.}(2017)\citenamefont {Martin},
  \citenamefont {Sayrafi},\ and\ \citenamefont {Whaley}}]{Martin_2017}%
  \BibitemOpen
  \bibfield  {author} {\bibinfo {author} {\bibfnamefont {L.}~\bibnamefont
  {Martin}}, \bibinfo {author} {\bibfnamefont {M.}~\bibnamefont {Sayrafi}}, \
  and\ \bibinfo {author} {\bibfnamefont {K.~B.}\ \bibnamefont {Whaley}},\
  }\href {\doibase 10.1088/2058-9565/aa804c} {\bibfield  {journal} {\bibinfo
  {journal} {Quantum Science and Technology}\ }\textbf {\bibinfo {volume}
  {2}},\ \bibinfo {pages} {044006} (\bibinfo {year} {2017})}\BibitemShut
  {NoStop}%
\bibitem [{\citenamefont {Cabrillo}\ \emph {et~al.}(1999)\citenamefont
  {Cabrillo}, \citenamefont {Cirac}, \citenamefont {Garc\'{\i}a-Fern\'andez},\
  and\ \citenamefont {Zoller}}]{Cabrillo1998}%
  \BibitemOpen
  \bibfield  {author} {\bibinfo {author} {\bibfnamefont {C.}~\bibnamefont
  {Cabrillo}}, \bibinfo {author} {\bibfnamefont {J.~I.}\ \bibnamefont {Cirac}},
  \bibinfo {author} {\bibfnamefont {P.}~\bibnamefont
  {Garc\'{\i}a-Fern\'andez}}, \ and\ \bibinfo {author} {\bibfnamefont
  {P.}~\bibnamefont {Zoller}},\ }\href {\doibase 10.1103/PhysRevA.59.1025}
  {\bibfield  {journal} {\bibinfo  {journal} {Phys. Rev. A}\ }\textbf {\bibinfo
  {volume} {59}},\ \bibinfo {pages} {1025} (\bibinfo {year}
  {1999})}\BibitemShut {NoStop}%
\bibitem [{\citenamefont {Plenio}\ \emph {et~al.}(1999)\citenamefont {Plenio},
  \citenamefont {Huelga}, \citenamefont {Beige},\ and\ \citenamefont
  {Knight}}]{Plenio1999}%
  \BibitemOpen
  \bibfield  {author} {\bibinfo {author} {\bibfnamefont {M.~B.}\ \bibnamefont
  {Plenio}}, \bibinfo {author} {\bibfnamefont {S.~F.}\ \bibnamefont {Huelga}},
  \bibinfo {author} {\bibfnamefont {A.}~\bibnamefont {Beige}}, \ and\ \bibinfo
  {author} {\bibfnamefont {P.~L.}\ \bibnamefont {Knight}},\ }\href {\doibase
  10.1103/PhysRevA.59.2468} {\bibfield  {journal} {\bibinfo  {journal} {Phys.
  Rev. A}\ }\textbf {\bibinfo {volume} {59}},\ \bibinfo {pages} {2468}
  (\bibinfo {year} {1999})}\BibitemShut {NoStop}%
\bibitem [{\citenamefont {Barrett}\ and\ \citenamefont
  {Kok}(2005)}]{BarrettKok}%
  \BibitemOpen
  \bibfield  {author} {\bibinfo {author} {\bibfnamefont {S.~D.}\ \bibnamefont
  {Barrett}}\ and\ \bibinfo {author} {\bibfnamefont {P.}~\bibnamefont {Kok}},\
  }\href {\doibase 10.1103/PhysRevA.71.060310} {\bibfield  {journal} {\bibinfo
  {journal} {Phys. Rev. A}\ }\textbf {\bibinfo {volume} {71}},\ \bibinfo
  {pages} {060310} (\bibinfo {year} {2005})}\BibitemShut {NoStop}%
\bibitem [{\citenamefont {Lim}\ \emph {et~al.}(2005)\citenamefont {Lim},
  \citenamefont {Beige},\ and\ \citenamefont {Kwek}}]{Lim2005}%
  \BibitemOpen
  \bibfield  {author} {\bibinfo {author} {\bibfnamefont {Y.~L.}\ \bibnamefont
  {Lim}}, \bibinfo {author} {\bibfnamefont {A.}~\bibnamefont {Beige}}, \ and\
  \bibinfo {author} {\bibfnamefont {L.~C.}\ \bibnamefont {Kwek}},\ }\href
  {\doibase 10.1103/PhysRevLett.95.030505} {\bibfield  {journal} {\bibinfo
  {journal} {Phys. Rev. Lett.}\ }\textbf {\bibinfo {volume} {95}},\ \bibinfo
  {pages} {030505} (\bibinfo {year} {2005})}\BibitemShut {NoStop}%
\bibitem [{\citenamefont {Moehring}\ \emph {et~al.}(2007)\citenamefont
  {Moehring}, \citenamefont {Maunz}, \citenamefont {Olmschenk}, \citenamefont
  {Younge}, \citenamefont {Matsukevich}, \citenamefont {Duan},\ and\
  \citenamefont {Monroe}}]{Moehring2007}%
  \BibitemOpen
  \bibfield  {author} {\bibinfo {author} {\bibfnamefont {D.~L.}\ \bibnamefont
  {Moehring}}, \bibinfo {author} {\bibfnamefont {P.}~\bibnamefont {Maunz}},
  \bibinfo {author} {\bibfnamefont {S.}~\bibnamefont {Olmschenk}}, \bibinfo
  {author} {\bibfnamefont {K.~C.}\ \bibnamefont {Younge}}, \bibinfo {author}
  {\bibfnamefont {D.~N.}\ \bibnamefont {Matsukevich}}, \bibinfo {author}
  {\bibfnamefont {L.-M.}\ \bibnamefont {Duan}}, \ and\ \bibinfo {author}
  {\bibfnamefont {C.}~\bibnamefont {Monroe}},\ }\href
  {https://www.nature.com/articles/nature06118} {\bibfield  {journal} {\bibinfo
   {journal} {Nature}\ }\textbf {\bibinfo {volume} {449}},\ \bibinfo {pages}
  {68} (\bibinfo {year} {2007})}\BibitemShut {NoStop}%
\bibitem [{\citenamefont {Maunz}\ \emph {et~al.}(2009)\citenamefont {Maunz},
  \citenamefont {Olmschenk}, \citenamefont {Hayes}, \citenamefont
  {Matsukevich}, \citenamefont {Duan},\ and\ \citenamefont
  {Monroe}}]{Maunz2009}%
  \BibitemOpen
  \bibfield  {author} {\bibinfo {author} {\bibfnamefont {P.}~\bibnamefont
  {Maunz}}, \bibinfo {author} {\bibfnamefont {S.}~\bibnamefont {Olmschenk}},
  \bibinfo {author} {\bibfnamefont {D.}~\bibnamefont {Hayes}}, \bibinfo
  {author} {\bibfnamefont {D.~N.}\ \bibnamefont {Matsukevich}}, \bibinfo
  {author} {\bibfnamefont {L.-M.}\ \bibnamefont {Duan}}, \ and\ \bibinfo
  {author} {\bibfnamefont {C.}~\bibnamefont {Monroe}},\ }\href
  {https://link.aps.org/doi/10.1103/PhysRevLett.102.250502} {\bibfield
  {journal} {\bibinfo  {journal} {Phys. Rev. Lett.}\ }\textbf {\bibinfo
  {volume} {102}},\ \bibinfo {pages} {250502} (\bibinfo {year}
  {2009})}\BibitemShut {NoStop}%
\bibitem [{\citenamefont {Hofmann}\ \emph {et~al.}(2012)\citenamefont
  {Hofmann}, \citenamefont {Krug}, \citenamefont {Ortegel}, \citenamefont
  {G{\'e}rard}, \citenamefont {Weber}, \citenamefont {Rosenfeld},\ and\
  \citenamefont {Weinfurter}}]{Hofmann2012}%
  \BibitemOpen
  \bibfield  {author} {\bibinfo {author} {\bibfnamefont {J.}~\bibnamefont
  {Hofmann}}, \bibinfo {author} {\bibfnamefont {M.}~\bibnamefont {Krug}},
  \bibinfo {author} {\bibfnamefont {N.}~\bibnamefont {Ortegel}}, \bibinfo
  {author} {\bibfnamefont {L.}~\bibnamefont {G{\'e}rard}}, \bibinfo {author}
  {\bibfnamefont {M.}~\bibnamefont {Weber}}, \bibinfo {author} {\bibfnamefont
  {W.}~\bibnamefont {Rosenfeld}}, \ and\ \bibinfo {author} {\bibfnamefont
  {H.}~\bibnamefont {Weinfurter}},\ }\href {\doibase 10.1126/science.1221856}
  {\bibfield  {journal} {\bibinfo  {journal} {Science}\ }\textbf {\bibinfo
  {volume} {337}},\ \bibinfo {pages} {72} (\bibinfo {year} {2012})}\BibitemShut
  {NoStop}%
\bibitem [{\citenamefont {Santos}\ \emph {et~al.}(2012)\citenamefont {Santos},
  \citenamefont {Terra~Cunha}, \citenamefont {Chaves},\ and\ \citenamefont
  {Carvalho}}]{Santos2012}%
  \BibitemOpen
  \bibfield  {author} {\bibinfo {author} {\bibfnamefont {M.~F.}\ \bibnamefont
  {Santos}}, \bibinfo {author} {\bibfnamefont {M.}~\bibnamefont {Terra~Cunha}},
  \bibinfo {author} {\bibfnamefont {R.}~\bibnamefont {Chaves}}, \ and\ \bibinfo
  {author} {\bibfnamefont {A.~R.~R.}\ \bibnamefont {Carvalho}},\ }\href
  {\doibase 10.1103/PhysRevLett.108.170501} {\bibfield  {journal} {\bibinfo
  {journal} {Phys. Rev. Lett.}\ }\textbf {\bibinfo {volume} {108}},\ \bibinfo
  {pages} {170501} (\bibinfo {year} {2012})}\BibitemShut {NoStop}%
\bibitem [{\citenamefont {Slodi\ifmmode~\check{c}\else \v{c}\fi{}ka}\ \emph
  {et~al.}(2013)\citenamefont {Slodi\ifmmode~\check{c}\else \v{c}\fi{}ka},
  \citenamefont {H\'etet}, \citenamefont {R\"ock}, \citenamefont {Schindler},
  \citenamefont {Hennrich},\ and\ \citenamefont {Blatt}}]{Slodika2013}%
  \BibitemOpen
  \bibfield  {author} {\bibinfo {author} {\bibfnamefont {L.}~\bibnamefont
  {Slodi\ifmmode~\check{c}\else \v{c}\fi{}ka}}, \bibinfo {author}
  {\bibfnamefont {G.}~\bibnamefont {H\'etet}}, \bibinfo {author} {\bibfnamefont
  {N.}~\bibnamefont {R\"ock}}, \bibinfo {author} {\bibfnamefont
  {P.}~\bibnamefont {Schindler}}, \bibinfo {author} {\bibfnamefont
  {M.}~\bibnamefont {Hennrich}}, \ and\ \bibinfo {author} {\bibfnamefont
  {R.}~\bibnamefont {Blatt}},\ }\href {\doibase 10.1103/PhysRevLett.110.083603}
  {\bibfield  {journal} {\bibinfo  {journal} {Phys. Rev. Lett.}\ }\textbf
  {\bibinfo {volume} {110}},\ \bibinfo {pages} {083603} (\bibinfo {year}
  {2013})}\BibitemShut {NoStop}%
\bibitem [{\citenamefont {Bernien}\ \emph {et~al.}(2013)\citenamefont
  {Bernien}, \citenamefont {Hensen}, \citenamefont {Pfaff}, \citenamefont
  {Koolstra}, \citenamefont {Blok}, \citenamefont {Robledo}, \citenamefont
  {Taminiau}, \citenamefont {Markham}, \citenamefont {Twitchen}, \citenamefont
  {Childress},\ and\ \citenamefont {Hanson}}]{Hanson2013Heralded}%
  \BibitemOpen
  \bibfield  {author} {\bibinfo {author} {\bibfnamefont {H.}~\bibnamefont
  {Bernien}}, \bibinfo {author} {\bibfnamefont {B.}~\bibnamefont {Hensen}},
  \bibinfo {author} {\bibfnamefont {W.}~\bibnamefont {Pfaff}}, \bibinfo
  {author} {\bibfnamefont {G.}~\bibnamefont {Koolstra}}, \bibinfo {author}
  {\bibfnamefont {M.~S.}\ \bibnamefont {Blok}}, \bibinfo {author}
  {\bibfnamefont {L.}~\bibnamefont {Robledo}}, \bibinfo {author} {\bibfnamefont
  {T.~H.}\ \bibnamefont {Taminiau}}, \bibinfo {author} {\bibfnamefont
  {M.}~\bibnamefont {Markham}}, \bibinfo {author} {\bibfnamefont {D.~J.}\
  \bibnamefont {Twitchen}}, \bibinfo {author} {\bibfnamefont {L.}~\bibnamefont
  {Childress}}, \ and\ \bibinfo {author} {\bibfnamefont {R.}~\bibnamefont
  {Hanson}},\ }\href {https://www.nature.com/articles/nature12016} {\bibfield
  {journal} {\bibinfo  {journal} {Nature}\ }\textbf {\bibinfo {volume} {497}},\
  \bibinfo {pages} {86} (\bibinfo {year} {2013})}\BibitemShut {NoStop}%
\bibitem [{\citenamefont {Pfaff}\ \emph {et~al.}(2014)\citenamefont {Pfaff},
  \citenamefont {Hensen}, \citenamefont {Bernien}, \citenamefont {van Dam},
  \citenamefont {Blok}, \citenamefont {Taminiau}, \citenamefont {Tiggelman},
  \citenamefont {Schouten}, \citenamefont {Markham}, \citenamefont {Twitchen},\
  and\ \citenamefont {Hanson}}]{Pfaff2014}%
  \BibitemOpen
  \bibfield  {author} {\bibinfo {author} {\bibfnamefont {W.}~\bibnamefont
  {Pfaff}}, \bibinfo {author} {\bibfnamefont {B.~J.}\ \bibnamefont {Hensen}},
  \bibinfo {author} {\bibfnamefont {H.}~\bibnamefont {Bernien}}, \bibinfo
  {author} {\bibfnamefont {S.~B.}\ \bibnamefont {van Dam}}, \bibinfo {author}
  {\bibfnamefont {M.~S.}\ \bibnamefont {Blok}}, \bibinfo {author}
  {\bibfnamefont {T.~H.}\ \bibnamefont {Taminiau}}, \bibinfo {author}
  {\bibfnamefont {M.~J.}\ \bibnamefont {Tiggelman}}, \bibinfo {author}
  {\bibfnamefont {R.~N.}\ \bibnamefont {Schouten}}, \bibinfo {author}
  {\bibfnamefont {M.}~\bibnamefont {Markham}}, \bibinfo {author} {\bibfnamefont
  {D.~J.}\ \bibnamefont {Twitchen}}, \ and\ \bibinfo {author} {\bibfnamefont
  {R.}~\bibnamefont {Hanson}},\ }\href
  {https://science.sciencemag.org/content/345/6196/532} {\bibfield  {journal}
  {\bibinfo  {journal} {Science}\ }\textbf {\bibinfo {volume} {345}},\ \bibinfo
  {pages} {532} (\bibinfo {year} {2014})}\BibitemShut {NoStop}%
\bibitem [{\citenamefont {Delteil}\ \emph {et~al.}(2016)\citenamefont
  {Delteil}, \citenamefont {Sun}, \citenamefont {Gao}, \citenamefont {Togan},
  \citenamefont {Faelt},\ and\ \citenamefont {Imamo\v{g}lu}}]{Delteil2016}%
  \BibitemOpen
  \bibfield  {author} {\bibinfo {author} {\bibfnamefont {A.}~\bibnamefont
  {Delteil}}, \bibinfo {author} {\bibfnamefont {Z.}~\bibnamefont {Sun}},
  \bibinfo {author} {\bibfnamefont {W.}~\bibnamefont {Gao}}, \bibinfo {author}
  {\bibfnamefont {E.}~\bibnamefont {Togan}}, \bibinfo {author} {\bibfnamefont
  {S.}~\bibnamefont {Faelt}}, \ and\ \bibinfo {author} {\bibfnamefont
  {A.}~\bibnamefont {Imamo\v{g}lu}},\ }\href
  {https://www.nature.com/articles/nphys3605?draft=collection} {\bibfield
  {journal} {\bibinfo  {journal} {Nat. Phys.}\ }\textbf {\bibinfo {volume}
  {12}},\ \bibinfo {pages} {218} (\bibinfo {year} {2016})}\BibitemShut
  {NoStop}%
\bibitem [{\citenamefont {Ohm}\ and\ \citenamefont {Hassler}(2017)}]{Ohm2017}%
  \BibitemOpen
  \bibfield  {author} {\bibinfo {author} {\bibfnamefont {C.}~\bibnamefont
  {Ohm}}\ and\ \bibinfo {author} {\bibfnamefont {F.}~\bibnamefont {Hassler}},\
  }\href {\doibase 10.1088/1367-2630/aa6d46} {\bibfield  {journal} {\bibinfo
  {journal} {New Journal of Physics}\ }\textbf {\bibinfo {volume} {19}},\
  \bibinfo {pages} {053018} (\bibinfo {year} {2017})}\BibitemShut {NoStop}%
\bibitem [{\citenamefont {Araneda}\ \emph {et~al.}(2018)\citenamefont
  {Araneda}, \citenamefont {Higginbottom}, \citenamefont
  {Slodi\ifmmode~\check{c}\else \v{c}\fi{}ka}, \citenamefont {Colombe},\ and\
  \citenamefont {Blatt}}]{Araneda2018}%
  \BibitemOpen
  \bibfield  {author} {\bibinfo {author} {\bibfnamefont {G.}~\bibnamefont
  {Araneda}}, \bibinfo {author} {\bibfnamefont {D.~B.}\ \bibnamefont
  {Higginbottom}}, \bibinfo {author} {\bibfnamefont {L.}~\bibnamefont
  {Slodi\ifmmode~\check{c}\else \v{c}\fi{}ka}}, \bibinfo {author}
  {\bibfnamefont {Y.}~\bibnamefont {Colombe}}, \ and\ \bibinfo {author}
  {\bibfnamefont {R.}~\bibnamefont {Blatt}},\ }\href {\doibase
  10.1103/PhysRevLett.120.193603} {\bibfield  {journal} {\bibinfo  {journal}
  {Phys. Rev. Lett.}\ }\textbf {\bibinfo {volume} {120}},\ \bibinfo {pages}
  {193603} (\bibinfo {year} {2018})}\BibitemShut {NoStop}%
\bibitem [{\citenamefont {Hensen}\ \emph {et~al.}(2015)\citenamefont {Hensen},
  \citenamefont {Bernien}, \citenamefont {Dr\'{e}au}, \citenamefont {Reiserer},
  \citenamefont {Kalb}, \citenamefont {Blok}, \citenamefont {Ruitenberg},
  \citenamefont {Vermeulen}, \citenamefont {Schouten}, \citenamefont
  {Abell\'{a}n}, \citenamefont {Amaya}, \citenamefont {Pruneri}, \citenamefont
  {Mitchell}, \citenamefont {Markham}, \citenamefont {Twitchen}, \citenamefont
  {Elkouss}, \citenamefont {Wehner}, \citenamefont {Taminiau},\ and\
  \citenamefont {Hanson}}]{HansonLoopholeFree}%
  \BibitemOpen
  \bibfield  {author} {\bibinfo {author} {\bibfnamefont {B.}~\bibnamefont
  {Hensen}}, \bibinfo {author} {\bibfnamefont {H.}~\bibnamefont {Bernien}},
  \bibinfo {author} {\bibfnamefont {A.~E.}\ \bibnamefont {Dr\'{e}au}}, \bibinfo
  {author} {\bibfnamefont {A.}~\bibnamefont {Reiserer}}, \bibinfo {author}
  {\bibfnamefont {N.}~\bibnamefont {Kalb}}, \bibinfo {author} {\bibfnamefont
  {M.~S.}\ \bibnamefont {Blok}}, \bibinfo {author} {\bibfnamefont
  {J.}~\bibnamefont {Ruitenberg}}, \bibinfo {author} {\bibfnamefont {R.~F.~L.}\
  \bibnamefont {Vermeulen}}, \bibinfo {author} {\bibfnamefont {R.~N.}\
  \bibnamefont {Schouten}}, \bibinfo {author} {\bibfnamefont {C.}~\bibnamefont
  {Abell\'{a}n}}, \bibinfo {author} {\bibfnamefont {W.}~\bibnamefont {Amaya}},
  \bibinfo {author} {\bibfnamefont {V.}~\bibnamefont {Pruneri}}, \bibinfo
  {author} {\bibfnamefont {M.~W.}\ \bibnamefont {Mitchell}}, \bibinfo {author}
  {\bibfnamefont {M.}~\bibnamefont {Markham}}, \bibinfo {author} {\bibfnamefont
  {D.~J.}\ \bibnamefont {Twitchen}}, \bibinfo {author} {\bibfnamefont
  {D.}~\bibnamefont {Elkouss}}, \bibinfo {author} {\bibfnamefont
  {S.}~\bibnamefont {Wehner}}, \bibinfo {author} {\bibfnamefont {T.~H.}\
  \bibnamefont {Taminiau}}, \ and\ \bibinfo {author} {\bibfnamefont
  {R.}~\bibnamefont {Hanson}},\ }\href
  {https://www.nature.com/articles/nature15759} {\bibfield  {journal} {\bibinfo
   {journal} {Nature}\ }\textbf {\bibinfo {volume} {526}},\ \bibinfo {pages}
  {682} (\bibinfo {year} {2015})}\BibitemShut {NoStop}%
\bibitem [{\citenamefont {Borregaard}\ \emph {et~al.}(2019)\citenamefont
  {Borregaard}, \citenamefont {S{\o}rensen},\ and\ \citenamefont
  {Lodahl}}]{Borregarrd2019}%
  \BibitemOpen
  \bibfield  {author} {\bibinfo {author} {\bibfnamefont {J.}~\bibnamefont
  {Borregaard}}, \bibinfo {author} {\bibfnamefont {A.~S.}\ \bibnamefont
  {S{\o}rensen}}, \ and\ \bibinfo {author} {\bibfnamefont {P.}~\bibnamefont
  {Lodahl}},\ }\href
  {https://onlinelibrary.wiley.com/doi/abs/10.1002/qute.201800091} {\bibfield
  {journal} {\bibinfo  {journal} {Advanced Quantum Technologies}\ }\textbf
  {\bibinfo {volume} {2}},\ \bibinfo {pages} {1800091} (\bibinfo {year}
  {2019})}\BibitemShut {NoStop}%
\bibitem [{\citenamefont {Yu}\ and\ \citenamefont
  {Eberly}(2004)}]{YuEberly_2004}%
  \BibitemOpen
  \bibfield  {author} {\bibinfo {author} {\bibfnamefont {T.}~\bibnamefont
  {Yu}}\ and\ \bibinfo {author} {\bibfnamefont {J.~H.}\ \bibnamefont
  {Eberly}},\ }\href {\doibase 10.1103/PhysRevLett.93.140404} {\bibfield
  {journal} {\bibinfo  {journal} {Phys. Rev. Lett.}\ }\textbf {\bibinfo
  {volume} {93}},\ \bibinfo {pages} {140404} (\bibinfo {year}
  {2004})}\BibitemShut {NoStop}%
\bibitem [{\citenamefont {Qian}\ and\ \citenamefont
  {Agarwal}(2019)}]{Xiao-Feng_2019}%
  \BibitemOpen
  \bibfield  {author} {\bibinfo {author} {\bibfnamefont {X.-F.}\ \bibnamefont
  {Qian}}\ and\ \bibinfo {author} {\bibfnamefont {G.~S.}\ \bibnamefont
  {Agarwal}},\ }\href {https://arxiv.org/abs/1901.07595} {\bibfield  {journal}
  {\bibinfo  {journal} {arXiv 1901.07595}\ } (\bibinfo {year}
  {2019})}\BibitemShut {NoStop}%
\bibitem [{\citenamefont {Campagne-Ibarcq}\ \emph {et~al.}(2014)\citenamefont
  {Campagne-Ibarcq}, \citenamefont {Bretheau}, \citenamefont {Flurin},
  \citenamefont {Auff\`eves}, \citenamefont {Mallet},\ and\ \citenamefont
  {Huard}}]{PCI-2013}%
  \BibitemOpen
  \bibfield  {author} {\bibinfo {author} {\bibfnamefont {P.}~\bibnamefont
  {Campagne-Ibarcq}}, \bibinfo {author} {\bibfnamefont {L.}~\bibnamefont
  {Bretheau}}, \bibinfo {author} {\bibfnamefont {E.}~\bibnamefont {Flurin}},
  \bibinfo {author} {\bibfnamefont {A.}~\bibnamefont {Auff\`eves}}, \bibinfo
  {author} {\bibfnamefont {F.}~\bibnamefont {Mallet}}, \ and\ \bibinfo {author}
  {\bibfnamefont {B.}~\bibnamefont {Huard}},\ }\href {\doibase
  10.1103/PhysRevLett.112.180402} {\bibfield  {journal} {\bibinfo  {journal}
  {Phys. Rev. Lett.}\ }\textbf {\bibinfo {volume} {112}},\ \bibinfo {pages}
  {180402} (\bibinfo {year} {2014})}\BibitemShut {NoStop}%
\bibitem [{\citenamefont {Jordan}\ \emph {et~al.}(2015)\citenamefont {Jordan},
  \citenamefont {Chantasri}, \citenamefont {Rouchon},\ and\ \citenamefont
  {Huard}}]{Jordan2015flor}%
  \BibitemOpen
  \bibfield  {author} {\bibinfo {author} {\bibfnamefont {A.~N.}\ \bibnamefont
  {Jordan}}, \bibinfo {author} {\bibfnamefont {A.}~\bibnamefont {Chantasri}},
  \bibinfo {author} {\bibfnamefont {P.}~\bibnamefont {Rouchon}}, \ and\
  \bibinfo {author} {\bibfnamefont {B.}~\bibnamefont {Huard}},\ }\href
  {\doibase 10.1007/s40509-016-0075-9} {\bibfield  {journal} {\bibinfo
  {journal} {Quantum Studies: Math. and Found.}\ }\textbf {\bibinfo {volume}
  {3}},\ \bibinfo {pages} {237} (\bibinfo {year} {2015})}\BibitemShut {NoStop}%
\bibitem [{\citenamefont {Campagne-Ibarcq}\ \emph
  {et~al.}(2016{\natexlab{a}})\citenamefont {Campagne-Ibarcq}, \citenamefont
  {Six}, \citenamefont {Bretheau}, \citenamefont {Sarlette}, \citenamefont
  {Mirrahimi}, \citenamefont {Rouchon},\ and\ \citenamefont
  {Huard}}]{Campagne-Ibarcq2016}%
  \BibitemOpen
  \bibfield  {author} {\bibinfo {author} {\bibfnamefont {P.}~\bibnamefont
  {Campagne-Ibarcq}}, \bibinfo {author} {\bibfnamefont {P.}~\bibnamefont
  {Six}}, \bibinfo {author} {\bibfnamefont {L.}~\bibnamefont {Bretheau}},
  \bibinfo {author} {\bibfnamefont {A.}~\bibnamefont {Sarlette}}, \bibinfo
  {author} {\bibfnamefont {M.}~\bibnamefont {Mirrahimi}}, \bibinfo {author}
  {\bibfnamefont {P.}~\bibnamefont {Rouchon}}, \ and\ \bibinfo {author}
  {\bibfnamefont {B.}~\bibnamefont {Huard}},\ }\href {\doibase
  10.1103/PhysRevX.6.011002} {\bibfield  {journal} {\bibinfo  {journal} {Phys.
  Rev. X}\ }\textbf {\bibinfo {volume} {6}},\ \bibinfo {pages} {011002}
  (\bibinfo {year} {2016}{\natexlab{a}})}\BibitemShut {NoStop}%
\bibitem [{\citenamefont {Campagne-Ibarcq}\ \emph
  {et~al.}(2016{\natexlab{b}})\citenamefont {Campagne-Ibarcq}, \citenamefont
  {Jezouin}, \citenamefont {Cottet}, \citenamefont {Six}, \citenamefont
  {Bretheau}, \citenamefont {Mallet}, \citenamefont {Sarlette}, \citenamefont
  {Rouchon},\ and\ \citenamefont {Huard}}]{PCI-2016-2}%
  \BibitemOpen
  \bibfield  {author} {\bibinfo {author} {\bibfnamefont {P.}~\bibnamefont
  {Campagne-Ibarcq}}, \bibinfo {author} {\bibfnamefont {S.}~\bibnamefont
  {Jezouin}}, \bibinfo {author} {\bibfnamefont {N.}~\bibnamefont {Cottet}},
  \bibinfo {author} {\bibfnamefont {P.}~\bibnamefont {Six}}, \bibinfo {author}
  {\bibfnamefont {L.}~\bibnamefont {Bretheau}}, \bibinfo {author}
  {\bibfnamefont {F.}~\bibnamefont {Mallet}}, \bibinfo {author} {\bibfnamefont
  {A.}~\bibnamefont {Sarlette}}, \bibinfo {author} {\bibfnamefont
  {P.}~\bibnamefont {Rouchon}}, \ and\ \bibinfo {author} {\bibfnamefont
  {B.}~\bibnamefont {Huard}},\ }\href {\doibase 10.1103/PhysRevLett.117.060502}
  {\bibfield  {journal} {\bibinfo  {journal} {Phys. Rev. Lett.}\ }\textbf
  {\bibinfo {volume} {117}},\ \bibinfo {pages} {060502} (\bibinfo {year}
  {2016}{\natexlab{b}})}\BibitemShut {NoStop}%
\bibitem [{\citenamefont {Naghiloo}\ \emph {et~al.}(2016)\citenamefont
  {Naghiloo}, \citenamefont {Foroozani}, \citenamefont {Tan}, \citenamefont
  {Jadbabaie},\ and\ \citenamefont {Murch}}]{Naghiloo2016flor}%
  \BibitemOpen
  \bibfield  {author} {\bibinfo {author} {\bibfnamefont {M.}~\bibnamefont
  {Naghiloo}}, \bibinfo {author} {\bibfnamefont {N.}~\bibnamefont {Foroozani}},
  \bibinfo {author} {\bibfnamefont {D.}~\bibnamefont {Tan}}, \bibinfo {author}
  {\bibfnamefont {A.}~\bibnamefont {Jadbabaie}}, \ and\ \bibinfo {author}
  {\bibfnamefont {K.~W.}\ \bibnamefont {Murch}},\ }\href {\doibase
  10.1038/ncomms11527} {\bibfield  {journal} {\bibinfo  {journal} {Nature
  Communications}\ }\textbf {\bibinfo {volume} {7}},\ \bibinfo {pages} {11527}
  (\bibinfo {year} {2016})}\BibitemShut {NoStop}%
\bibitem [{\citenamefont {Naghiloo}\ \emph {et~al.}(2017)\citenamefont
  {Naghiloo}, \citenamefont {Tan}, \citenamefont {Harrington}, \citenamefont
  {Lewalle}, \citenamefont {Jordan},\ and\ \citenamefont {Murch}}]{Mahdi2016}%
  \BibitemOpen
  \bibfield  {author} {\bibinfo {author} {\bibfnamefont {M.}~\bibnamefont
  {Naghiloo}}, \bibinfo {author} {\bibfnamefont {D.}~\bibnamefont {Tan}},
  \bibinfo {author} {\bibfnamefont {P.~M.}\ \bibnamefont {Harrington}},
  \bibinfo {author} {\bibfnamefont {P.}~\bibnamefont {Lewalle}}, \bibinfo
  {author} {\bibfnamefont {A.~N.}\ \bibnamefont {Jordan}}, \ and\ \bibinfo
  {author} {\bibfnamefont {K.~W.}\ \bibnamefont {Murch}},\ }\href {\doibase
  10.1103/PhysRevA.96.053807} {\bibfield  {journal} {\bibinfo  {journal} {Phys.
  Rev. A}\ }\textbf {\bibinfo {volume} {96}},\ \bibinfo {pages} {053807}
  (\bibinfo {year} {2017})}\BibitemShut {NoStop}%
\bibitem [{\citenamefont {Tan}\ \emph {et~al.}(2017)\citenamefont {Tan},
  \citenamefont {Foroozani}, \citenamefont {Naghiloo}, \citenamefont
  {Kiilerich}, \citenamefont {M\o{}lmer},\ and\ \citenamefont
  {Murch}}]{Tan2017}%
  \BibitemOpen
  \bibfield  {author} {\bibinfo {author} {\bibfnamefont {D.}~\bibnamefont
  {Tan}}, \bibinfo {author} {\bibfnamefont {N.}~\bibnamefont {Foroozani}},
  \bibinfo {author} {\bibfnamefont {M.}~\bibnamefont {Naghiloo}}, \bibinfo
  {author} {\bibfnamefont {A.~H.}\ \bibnamefont {Kiilerich}}, \bibinfo {author}
  {\bibfnamefont {K.}~\bibnamefont {M\o{}lmer}}, \ and\ \bibinfo {author}
  {\bibfnamefont {K.~W.}\ \bibnamefont {Murch}},\ }\href {\doibase
  10.1103/PhysRevA.96.022104} {\bibfield  {journal} {\bibinfo  {journal} {Phys.
  Rev. A}\ }\textbf {\bibinfo {volume} {96}},\ \bibinfo {pages} {022104}
  (\bibinfo {year} {2017})}\BibitemShut {NoStop}%
\bibitem [{\citenamefont {Ficheux}\ \emph {et~al.}(2018)\citenamefont
  {Ficheux}, \citenamefont {Jezouin}, \citenamefont {Leghtas},\ and\
  \citenamefont {Huard}}]{Ficheux2018}%
  \BibitemOpen
  \bibfield  {author} {\bibinfo {author} {\bibfnamefont {Q.}~\bibnamefont
  {Ficheux}}, \bibinfo {author} {\bibfnamefont {S.}~\bibnamefont {Jezouin}},
  \bibinfo {author} {\bibfnamefont {Z.}~\bibnamefont {Leghtas}}, \ and\
  \bibinfo {author} {\bibfnamefont {B.}~\bibnamefont {Huard}},\ }\href
  {https://www.nature.com/articles/s41467-018-04372-9} {\bibfield  {journal}
  {\bibinfo  {journal} {Nat. Comm.}\ }\textbf {\bibinfo {volume} {9}},\
  \bibinfo {pages} {1926} (\bibinfo {year} {2018})}\BibitemShut {NoStop}%
\bibitem [{\citenamefont {Lewalle}\ \emph
  {et~al.}(2019{\natexlab{a}})\citenamefont {Lewalle}, \citenamefont
  {Manikandan}, \citenamefont {Elouard},\ and\ \citenamefont
  {Jordan}}]{FlorTeach2019}%
  \BibitemOpen
  \bibfield  {author} {\bibinfo {author} {\bibfnamefont {P.}~\bibnamefont
  {Lewalle}}, \bibinfo {author} {\bibfnamefont {S.~K.}\ \bibnamefont
  {Manikandan}}, \bibinfo {author} {\bibfnamefont {C.}~\bibnamefont {Elouard}},
  \ and\ \bibinfo {author} {\bibfnamefont {A.~N.}\ \bibnamefont {Jordan}},\
  }\href {https://arxiv.org/abs/1908.04720v2} {\bibfield  {journal} {\bibinfo
  {journal} {arxiv 1908.04720v2}\ } (\bibinfo {year}
  {2019}{\natexlab{a}})}\BibitemShut {NoStop}%
\bibitem [{\citenamefont {Carvalho}\ \emph {et~al.}(2007)\citenamefont
  {Carvalho}, \citenamefont {Busse}, \citenamefont {Brodier}, \citenamefont
  {Viviescas},\ and\ \citenamefont {Buchleitner}}]{Carvalho2007}%
  \BibitemOpen
  \bibfield  {author} {\bibinfo {author} {\bibfnamefont {A.~R.~R.}\
  \bibnamefont {Carvalho}}, \bibinfo {author} {\bibfnamefont {M.}~\bibnamefont
  {Busse}}, \bibinfo {author} {\bibfnamefont {O.}~\bibnamefont {Brodier}},
  \bibinfo {author} {\bibfnamefont {C.}~\bibnamefont {Viviescas}}, \ and\
  \bibinfo {author} {\bibfnamefont {A.}~\bibnamefont {Buchleitner}},\ }\href
  {https://link.aps.org/doi/10.1103/PhysRevLett.98.190501} {\bibfield
  {journal} {\bibinfo  {journal} {Phys. Rev. Lett.}\ }\textbf {\bibinfo
  {volume} {98}},\ \bibinfo {pages} {190501} (\bibinfo {year}
  {2007})}\BibitemShut {NoStop}%
\bibitem [{\citenamefont {Viviescas}\ \emph {et~al.}(2010)\citenamefont
  {Viviescas}, \citenamefont {Guevara}, \citenamefont {Carvalho}, \citenamefont
  {Busse},\ and\ \citenamefont {Buchleitner}}]{Viviescas2010}%
  \BibitemOpen
  \bibfield  {author} {\bibinfo {author} {\bibfnamefont {C.}~\bibnamefont
  {Viviescas}}, \bibinfo {author} {\bibfnamefont {I.}~\bibnamefont {Guevara}},
  \bibinfo {author} {\bibfnamefont {A.~R.~R.}\ \bibnamefont {Carvalho}},
  \bibinfo {author} {\bibfnamefont {M.}~\bibnamefont {Busse}}, \ and\ \bibinfo
  {author} {\bibfnamefont {A.}~\bibnamefont {Buchleitner}},\ }\href {\doibase
  10.1103/PhysRevLett.105.210502} {\bibfield  {journal} {\bibinfo  {journal}
  {Phys. Rev. Lett.}\ }\textbf {\bibinfo {volume} {105}},\ \bibinfo {pages}
  {210502} (\bibinfo {year} {2010})}\BibitemShut {NoStop}%
\bibitem [{\citenamefont {Mascarenhas}\ \emph {et~al.}(2011)\citenamefont
  {Mascarenhas}, \citenamefont {Cavalcanti}, \citenamefont {Vedral},\ and\
  \citenamefont {Santos}}]{Mascarenhas2011}%
  \BibitemOpen
  \bibfield  {author} {\bibinfo {author} {\bibfnamefont {E.}~\bibnamefont
  {Mascarenhas}}, \bibinfo {author} {\bibfnamefont {D.}~\bibnamefont
  {Cavalcanti}}, \bibinfo {author} {\bibfnamefont {V.}~\bibnamefont {Vedral}},
  \ and\ \bibinfo {author} {\bibfnamefont {M.~F.}\ \bibnamefont {Santos}},\
  }\href {\doibase 10.1103/PhysRevA.83.022311} {\bibfield  {journal} {\bibinfo
  {journal} {Phys. Rev. A}\ }\textbf {\bibinfo {volume} {83}},\ \bibinfo
  {pages} {022311} (\bibinfo {year} {2011})}\BibitemShut {NoStop}%
\bibitem [{\citenamefont {Mintert}\ \emph {et~al.}(2005)\citenamefont
  {Mintert}, \citenamefont {Carvalho}, \citenamefont {Ku\'{s}},\ and\
  \citenamefont {Buchleitner}}]{Mintert2005}%
  \BibitemOpen
  \bibfield  {author} {\bibinfo {author} {\bibfnamefont {F.}~\bibnamefont
  {Mintert}}, \bibinfo {author} {\bibfnamefont {A.~R.~R.}\ \bibnamefont
  {Carvalho}}, \bibinfo {author} {\bibfnamefont {M.}~\bibnamefont {Ku\'{s}}}, \
  and\ \bibinfo {author} {\bibfnamefont {A.}~\bibnamefont {Buchleitner}},\
  }\href {\doibase https://doi.org/10.1016/j.physrep.2005.04.006} {\bibfield
  {journal} {\bibinfo  {journal} {Physics Reports}\ }\textbf {\bibinfo {volume}
  {415}},\ \bibinfo {pages} {207 } (\bibinfo {year} {2005})}\BibitemShut
  {NoStop}%
\bibitem [{\citenamefont {Mascarenhas}\ \emph
  {et~al.}(2010{\natexlab{a}})\citenamefont {Mascarenhas}, \citenamefont
  {Marques}, \citenamefont {Cunha},\ and\ \citenamefont
  {Santos}}]{Mascarenhas2010}%
  \BibitemOpen
  \bibfield  {author} {\bibinfo {author} {\bibfnamefont {E.}~\bibnamefont
  {Mascarenhas}}, \bibinfo {author} {\bibfnamefont {B.}~\bibnamefont
  {Marques}}, \bibinfo {author} {\bibfnamefont {M.~T.}\ \bibnamefont {Cunha}},
  \ and\ \bibinfo {author} {\bibfnamefont {M.~F.}\ \bibnamefont {Santos}},\
  }\href {\doibase 10.1103/PhysRevA.82.032327} {\bibfield  {journal} {\bibinfo
  {journal} {Phys. Rev. A}\ }\textbf {\bibinfo {volume} {82}},\ \bibinfo
  {pages} {032327} (\bibinfo {year} {2010}{\natexlab{a}})}\BibitemShut
  {NoStop}%
\bibitem [{\citenamefont {Carvalho}\ and\ \citenamefont
  {Santos}(2011)}]{Carvalho2011}%
  \BibitemOpen
  \bibfield  {author} {\bibinfo {author} {\bibfnamefont {A.~R.~R.}\
  \bibnamefont {Carvalho}}\ and\ \bibinfo {author} {\bibfnamefont {M.~F.}\
  \bibnamefont {Santos}},\ }\href {\doibase 10.1088/1367-2630/13/1/013010}
  {\bibfield  {journal} {\bibinfo  {journal} {New Journal of Physics}\ }\textbf
  {\bibinfo {volume} {13}},\ \bibinfo {pages} {013010} (\bibinfo {year}
  {2011})}\BibitemShut {NoStop}%
\bibitem [{\citenamefont {Nielsen}\ and\ \citenamefont
  {Chuang}(2000)}]{BookNielsen}%
  \BibitemOpen
  \bibfield  {author} {\bibinfo {author} {\bibfnamefont {M.~A.}\ \bibnamefont
  {Nielsen}}\ and\ \bibinfo {author} {\bibfnamefont {I.~L.}\ \bibnamefont
  {Chuang}},\ }\href@noop {} {\emph {\bibinfo {title} {Quantum Computation and
  Quantum Information}}}\ (\bibinfo  {publisher} {Cambridge University Press},\
  \bibinfo {year} {2000})\BibitemShut {NoStop}%
\bibitem [{\citenamefont {Lewalle}\ \emph
  {et~al.}(2019{\natexlab{b}})\citenamefont {Lewalle}, \citenamefont {Elouard},
  \citenamefont {Manikandan}, \citenamefont {Qian}, \citenamefont {Eberly},\
  and\ \citenamefont {Jordan}}]{LongFlor2019}%
  \BibitemOpen
  \bibfield  {author} {\bibinfo {author} {\bibfnamefont {P.}~\bibnamefont
  {Lewalle}}, \bibinfo {author} {\bibfnamefont {C.}~\bibnamefont {Elouard}},
  \bibinfo {author} {\bibfnamefont {S.~K.}\ \bibnamefont {Manikandan}},
  \bibinfo {author} {\bibfnamefont {X.-F.}\ \bibnamefont {Qian}}, \bibinfo
  {author} {\bibfnamefont {J.~H.}\ \bibnamefont {Eberly}}, \ and\ \bibinfo
  {author} {\bibfnamefont {A.~N.}\ \bibnamefont {Jordan}},\ }\href
  {https://arxiv.org/abs/1910.01206} {\bibfield  {journal} {\bibinfo  {journal}
  {arXiv:1910.01206}\ } (\bibinfo {year} {2019}{\natexlab{b}})}\BibitemShut
  {NoStop}%
\bibitem [{\citenamefont {Einstein}\ \emph {et~al.}(1935)\citenamefont
  {Einstein}, \citenamefont {Podolsky},\ and\ \citenamefont {Rosen}}]{EPR1935}%
  \BibitemOpen
  \bibfield  {author} {\bibinfo {author} {\bibfnamefont {A.}~\bibnamefont
  {Einstein}}, \bibinfo {author} {\bibfnamefont {B.}~\bibnamefont {Podolsky}},
  \ and\ \bibinfo {author} {\bibfnamefont {N.}~\bibnamefont {Rosen}},\ }\href
  {https://link.aps.org/doi/10.1103/PhysRev.47.777} {\bibfield  {journal}
  {\bibinfo  {journal} {Phys. Rev.}\ }\textbf {\bibinfo {volume} {47}},\
  \bibinfo {pages} {777} (\bibinfo {year} {1935})}\BibitemShut {NoStop}%
\bibitem [{\citenamefont {Takeda}\ \emph {et~al.}(2015)\citenamefont {Takeda},
  \citenamefont {Fuwa}, \citenamefont {van Loock},\ and\ \citenamefont
  {Furusawa}}]{Takeda2015}%
  \BibitemOpen
  \bibfield  {author} {\bibinfo {author} {\bibfnamefont {S.}~\bibnamefont
  {Takeda}}, \bibinfo {author} {\bibfnamefont {M.}~\bibnamefont {Fuwa}},
  \bibinfo {author} {\bibfnamefont {P.}~\bibnamefont {van Loock}}, \ and\
  \bibinfo {author} {\bibfnamefont {A.}~\bibnamefont {Furusawa}},\ }\href
  {\doibase 10.1103/PhysRevLett.114.100501} {\bibfield  {journal} {\bibinfo
  {journal} {Phys. Rev. Lett.}\ }\textbf {\bibinfo {volume} {114}},\ \bibinfo
  {pages} {100501} (\bibinfo {year} {2015})}\BibitemShut {NoStop}%
\bibitem [{\citenamefont {Flurin}\ \emph {et~al.}(2012)\citenamefont {Flurin},
  \citenamefont {Roch}, \citenamefont {Mallet}, \citenamefont {Devoret},\ and\
  \citenamefont {Huard}}]{Flurin2012}%
  \BibitemOpen
  \bibfield  {author} {\bibinfo {author} {\bibfnamefont {E.}~\bibnamefont
  {Flurin}}, \bibinfo {author} {\bibfnamefont {N.}~\bibnamefont {Roch}},
  \bibinfo {author} {\bibfnamefont {F.}~\bibnamefont {Mallet}}, \bibinfo
  {author} {\bibfnamefont {M.~H.}\ \bibnamefont {Devoret}}, \ and\ \bibinfo
  {author} {\bibfnamefont {B.}~\bibnamefont {Huard}},\ }\href {\doibase
  10.1103/PhysRevLett.109.183901} {\bibfield  {journal} {\bibinfo  {journal}
  {Phys. Rev. Lett.}\ }\textbf {\bibinfo {volume} {109}},\ \bibinfo {pages}
  {183901} (\bibinfo {year} {2012})}\BibitemShut {NoStop}%
\bibitem [{\citenamefont {Wootters}(1998)}]{wooters1998}%
  \BibitemOpen
  \bibfield  {author} {\bibinfo {author} {\bibfnamefont {W.~K.}\ \bibnamefont
  {Wootters}},\ }\href {\doibase 10.1103/PhysRevLett.80.2245} {\bibfield
  {journal} {\bibinfo  {journal} {Phys. Rev. Lett.}\ }\textbf {\bibinfo
  {volume} {80}},\ \bibinfo {pages} {2245} (\bibinfo {year}
  {1998})}\BibitemShut {NoStop}%
\bibitem [{\citenamefont {Mascarenhas}\ \emph
  {et~al.}(2010{\natexlab{b}})\citenamefont {Mascarenhas}, \citenamefont
  {Marques}, \citenamefont {Cavalcanti}, \citenamefont {Cunha},\ and\
  \citenamefont {Santos}}]{Mascarenhas2010-1}%
  \BibitemOpen
  \bibfield  {author} {\bibinfo {author} {\bibfnamefont {E.}~\bibnamefont
  {Mascarenhas}}, \bibinfo {author} {\bibfnamefont {B.}~\bibnamefont
  {Marques}}, \bibinfo {author} {\bibfnamefont {D.}~\bibnamefont {Cavalcanti}},
  \bibinfo {author} {\bibfnamefont {M.~T.}\ \bibnamefont {Cunha}}, \ and\
  \bibinfo {author} {\bibfnamefont {M.~F.}\ \bibnamefont {Santos}},\ }\href
  {\doibase 10.1103/PhysRevA.81.032310} {\bibfield  {journal} {\bibinfo
  {journal} {Phys. Rev. A}\ }\textbf {\bibinfo {volume} {81}},\ \bibinfo
  {pages} {032310} (\bibinfo {year} {2010}{\natexlab{b}})}\BibitemShut
  {NoStop}%
\bibitem [{\citenamefont {Martin}\ and\ \citenamefont
  {Whaley}(2019)}]{Leigh2019}%
  \BibitemOpen
  \bibfield  {author} {\bibinfo {author} {\bibfnamefont {L.~S.}\ \bibnamefont
  {Martin}}\ and\ \bibinfo {author} {\bibfnamefont {K.~B.}\ \bibnamefont
  {Whaley}},\ }\href {https://arxiv.org/abs/1912.00067} {\bibfield  {journal}
  {\bibinfo  {journal} {arXiv:1912.00067}\ } (\bibinfo {year}
  {2019})}\BibitemShut {NoStop}%
\end{thebibliography}%
\end{document}